\documentclass[%
aps,
notitlepage,
amsmath,amssymb,
reprint,%
longbibliography,
floatfix,
nofootinbib,
superscriptaddress]{revtex4-2}
\usepackage{graphicx} 
\usepackage{hyperref}
\hypersetup{colorlinks=true,
	linkcolor=black,
	allcolors=black}
\usepackage[utf8]{inputenc}
\usepackage{upgreek}
\usepackage[version=4]{mhchem}
\usepackage{physics}
\usepackage{siunitx}

\newcommand{\coschi}{\cos(\chi)}
\usepackage{float}
\AtBeginDocument{\RenewCommandCopy\qty\SI}
\usepackage{xcolor}
\newcommand{\highlight}[1]{{#1}}
\newcommand{\hl}[1]{\highlight{#1}}

\makeatletter
\begin{document}

\begin{titlepage}
\title{Enantioselective radical reactions can be induced by electron spin polarization: A quantum mechanism for Nature's emergent homochirality?}
\author{Thomas P. Fay}
\email[Corresponding author:\ ]{thomaspfay@ucla.edu}
\affiliation{Department of Chemistry and Biochemistry, University of California, Los Angeles, California 90095, United States}



\begin{abstract}
    Biomolecules that constitute life on Earth are chiral, but the precise mechanism by which homochirality emerged remains a mystery. In this work it is demonstrated that reactions of radical pairs, where one of the radical electron spins is polarised, can be enantioselective. This phenomenon arises from transient coherent quantum dynamics of the radical pair electron spins, which is known to occur even in warm and noisy condensed phase environments, where energetic perturbations much smaller than thermal energy can have large effects on reactivity. A quantitative theory is presented based on the molecular theory of chirality induced spin selectivity (CISS), where electron exchange interactions and chirality-dependent spin-orbit coupling effects control enantioselectivity. This theory provides useful bounds on the maximum enantiomeric excess for these reactions, which are found to be consistent with previous experiments. The enantioseletive radical pair mechanism presented here provides an alternative mechanistic basis to a recent proposal that spin-polarised photoelectrons from magnetite provided the initial chiral symmetry breaking necessary for the inception of homochirality in Nature, as well as suggests a new strategy for asymmetric synthesis using spin-polarised electrons.
\end{abstract}

\maketitle
\end{titlepage}

All bio-polymers in Nature, including proteins and DNA, are chiral, meaning they cannot be superimposed on their mirror image.\cite{blackmond_origin_2019,sallembien_possible_2022} The emergence of this homochirality is intriguing, given that in isolation the two mirror images (enantiomers) of a chiral molecule are energetically equivalent, so entropy should strongly favour an equal mixture of enantiomers in Nature, in contradiction to what is observed. Therefore understanding the emergent homochirality in Nature is considered essential for understanding the origin of life.\cite{blackmond_origin_2019,sallembien_possible_2022}

Nature's homochirality is generally thought to depend on two processes occurring in the prebiotic world: firstly a chiral symmetry breaking process which produces an excess of one enantiomer over another, and secondly a mechanism of amplification of this enantiomeric excess.\cite{blackmond_origin_2019,sallembien_possible_2022} Several specific mechanisms have been explored for the latter amplification process, which generally rely of some non-linear kinetics of chemical reactions involving chiral reactants. This work however focuses on the first step of chiral symmetry breaking.\cite{blackmond_origin_2019} Many mechanisms for chiral symmetry breaking have been proposed including circularly polarised light,\cite{rau_asymmetric_1983} magnetic fields,\cite{rikken_enantioselective_2000} and the chirality induced spin-selectivity (CISS) effect,\cite{ozturk_origins_2022,ozturk_origin_2023,ozturk_central_2023} among others.\cite{blackmond_origin_2019,sallembien_possible_2022} The recent experiments of Metzger \textit{et al.},\cite{metzger_electron_2020,metzger_dynamic_2021} as well as other researchers,\cite{bhowmick_spin-induced_2022} have shown that spin-polarised solid surface catalysts can asymmetrically catalyse \hl{organic} reactions, yielding a modest enantiomeric excess in the final products \hl{of between 8.5\% and 16\%},\cite{metzger_dynamic_2021} which has been at least qualitatively explained through the CISS effect.\cite{bloom_chiral_2024} This lends some support to the general proposal that CISS could effect the initial chiral symmetry breaking necessary for the emergence of homochirality in nature. Furthermore Subotnik and co-workers\cite{wu_electronic_2021,bian_spin-dependent_2024} have shown theoretically that reactive scattering can depend on electron spin polarisation. \hl{To date however there is only limited little experimental evidence for spin-polarization leading to enantioselective reactions,\cite{metzger_electron_2020,metzger_dynamic_2021} perhaps because there is no theory directly applicable to condensed phase reactions for this effect.}

Ozturk and Sasselov suggested that photoelectrons emitted from magnetised magnetite in prebiotic lakes could produce chiral symmetry breaking through the CISS effect.\cite{ozturk_origins_2022} They proposed that energy splitting between states of chiral molecules interacting with spin-polarised photo-electrons gives rise to a difference in reaction barriers for reduction of different enantiomers, which would give rise to an enantiomeric excess in reaction products from a racemic mixture (50:50 mixture of reactant enantiomers),\cite{ozturk_origins_2022} although more recently Ozturk \& co-workers have also explored other possibilities, namely preferential adsorption of chiral molecules onto magnetized surfaces.\cite{ozturk_origin_2023,ozturk_chirality-induced_2023} 
As an alternative to this proposal, in this letter a detailed microscopic mechanism of electron spin-induced chiral selectivity, where the key step is a redox reaction between radicals, one of which is spin-polarized. This model applies generally to spin-polarised radical recombination reactions, which could form the basis of how homochirality emerged in the prebiotic world, and could also form the basis of a new strategy for asymmetric synthesis. This enantioselective radical pair mechanism involves well-studied coherent quantum dynamics of electron spins in radicals in the condensed phase, where spin-polarized radicals could be formed by electrons from magnetized materials transferring to chemical fragments on their surface. Radical pair reactions are well known to be sensitive to weak magnetic perturbations to the electron spin dynamics,\cite{steiner_magnetic_1989,mani_molecular_2022} even magnetic fields as weak as 50 $\upmu$T,\cite{maeda_chemical_2008} as has been suggested to play a role in the magnetic field sense of migratory birds,\cite{hore_radical-pair_2016} so it stands to reason that radical pair reactions can also be sensitive to small perturbations from chirality-dependent spin-orbit coupling.\cite{bloom_chiral_2024} This idea is expanded upon below, where a qualitative description of the mechanism is given, which is then expanded into a more complete Stochastic Liouville equation model. This model is then used to derive simple analytic expressions for the enantiomeric excess in terms of kinetic and spin-interaction parameters for the radical pair. \hl{We discuss an example chemical reaction which could show this form of spin-polarisation induced enantioselectivity: the reduction of glyceronitrile at magnetite surfaces, which could have created an enantiomeric imbalance in prebiotic chemistry.}
\begin{figure}[t]
    \centering
    \includegraphics[width=\linewidth]{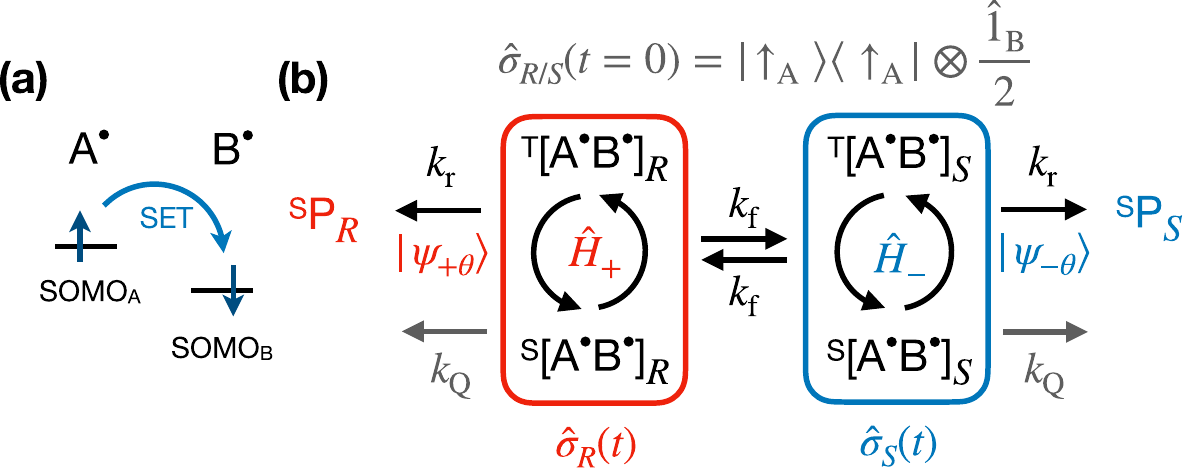}
    {\footnotesize\caption{(a) An illustration of the frontier orbitals of the radical pair \ce{[A^$\bullet $B^$\bullet$]} we consider here, and the SET reaction that is spin and chirality selective. (b) An illustration of the proposed mechanism for spin-polarisation induced enantioselective chemical reactions. Straight arrows indicate chemical reactions and curved arrows represent coherence quantum dynamics, with the red and blue arrows indicating the different coherent dynamics of the $R$ and $S$ intermediate radical pairs.}\label{fig-scheme}}
    \vspace{0pt}
\end{figure}

Experimental\cite{carmeli_spin_2014,eckvahl_direct_2023,eckvahl_detecting_2024} and theoretical works\cite{fay_chirality-induced_2021,fay_origin_2021,fay_spin_2023,chandran_electron_2022,chandran_effect_2022,chiesa_many-body_2024} have found that electron transfer in chiral radical pair systems can be spin-polarisation selective. This chirality induced spin selectivity effect is understood as emerging from spin-symmetry breaking in chiral molecules arising from spin-orbit coupling,\cite{bloom_chiral_2024,fransson_chiral_2025,chiesa_chirality-induced_2025} and energy dissipation and decoherence induced by the thermal environment.\cite{fay_origin_2021,fransson_chiral_2025} It therefore seems reasonable to ask, can reactions of spin-polarised radicals be chirality selective? The key intermediate that will be considered in the potential spin-control of enantioselective reactions is a pair of radicals (or radical ions) \ce{A^$\bullet$} and \ce{B^$\bullet$}. It is assumed that the \ce{[A^$\bullet $B^$\bullet$]} system is prochiral, and there exist two rapidly converting chiral forms of the \ce{A^$\bullet $B^$\bullet$} system, which are denoted \ce{[A^$\bullet $B^$\bullet$]_{$R/S$}},\hl{ which react irreversibly to form a chiral product}. These two forms have the same free energy, so they are initially equally populated. The radical pair undergoes a spin selective electron (or hole) transfer to a singlet product \ce{{}^SP_{$R/S$}}, and after this recombination the $R/S$ enantiomers cannot interconvert. The \ce{{}^SP_{$R/S$}} species could correspond to the final chiral product or could itself be an intermediate which forms some other final products. 
Here it is assumed that the key step involves electron transfer but hole transfer could equally control this process and the theory presented applies equally in this case. This scheme is fairly general, for example \ce{A^{$\bullet$}} could could correspond to a partially oxidised (or reduced) intermediate which fluxionally interconverts between enantiomers, and \ce{B^{$\bullet$}} to an oxidising agent (or reducing agent) which completes the reaction and fixes the chirality of the final product. \hl{The important steps in the reaction scheme can be summarised as
\begin{eqnarray*}
\ce{[A^$\bullet $B^$\bullet$]_{$R$} <=>[{$k_\mathrm{f}$}][{$k_\mathrm{f}$}] [A^$\bullet $B^$\bullet$]_{$S$}} \\
\ce{[A^$\bullet $B^$\bullet$]_{$R$} ->[{$k_\mathrm{r}$}] {}^SP_{$R$} }\\
\ce{[A^$\bullet $B^$\bullet$]_{$S$} ->[{$k_\mathrm{r}$}] {}^SP_{$S$} }.
\end{eqnarray*}}
Importantly, quantum coherence between radical pair singlet and triplet spin states can be long-lived, as is well established both experimentally and theoretically, so \ce{[A^$\bullet $B^$\bullet$]_{$R/S$}} will undergo coherent spin dynamics; this is illustrated in Fig.~\ref{fig-scheme}. This coherence is relatively robust to decoherence,\cite{steiner_magnetic_1989} and it can exist naturally for microseconds in condensed phase,\cite{maeda_chemical_2008,mims_readout_2021,malrieu_magnetic_2014} room temperature systems, which is plenty of time for coherent dynamics to influence chemical reactivity.

Previous work on electron transfer in chiral systems has shown that when spin-orbit coupling mediates an electron transfer in a chiral system, it rotates the spin of the electron that is being transferred.\cite{fay_chirality-induced_2021} \hl{Here we consider a single electron transfer reaction from a chiral \ce{A^$\bullet$B^$\bullet$} radical (ion) pair state, to a closed shell singlet state product \ce{${}^\mathrm{S}$P}, where an electron transfers from the singly occupied molecular orbital (SOMO) on A to the SOMO on B (as illustrated in Fig.~\ref{fig-scheme}(a)).} \hl{This electron transfer occurs via both direct spin-conserving electronic coupling, $V_\mathrm{AB}$ between the SOMOs, and a spin-orbit charge transfer coupling, $\Lambda_{\mathrm{AB}}$ between the SOMOs. Accounting for these two coupling mechanisms, the singlet $\mathrm{S}$ and triplet $\mathrm{T}_z = \mathrm{T}_0$ \ce{A^$\bullet $B^$\bullet$} states and final $\ce{{}^SP}$ electron transfer states are coupled by the electron transfer Hamiltonian $\hat{H}_\mathrm{ET}$}
\begin{align}
\hat{H}_{\mathrm{ET}} &= V_\mathrm{AB}\dyad{\ce{{}^SP}}{\ce{{}^S[A^$\bullet $B^$\bullet$}]} + i \frac{\Lambda_{\mathrm{AB}}}{2} \dyad{\ce{{}^SP}}{\ce{{}^{T_0}[A^$\bullet $B^$\bullet$}]}+ \mathrm{h.c.}\nonumber\\
&= \dyad{\ce{{}^SP}}{\ce{[A^$\bullet $B^$\bullet$]}}\hat{P}_\mathrm{S}(V_{\mathrm{AB}} + i {\Lambda}_{\mathrm{AB}}\hat{{S}}_{\mathrm{A}z}) + \mathrm{h.c.}
\end{align}
where $\hat{P}_\mathrm{S}$ is a projection operator onto the singlet spin states of the radical pair \ce{A^$\bullet $B^$\bullet$} and $\hat{{S}}_{\mathrm{A}z}$ is the $z$ component of the spin operator for the electron being transferredxxx. 
The rotation axis is determined by the direction of the spin-orbit coupling vector $\vb*{\Lambda}_{\ce{AB}}$ that couples the orbitals in \ce{A^{$\bullet$}} and \ce{B^{$\bullet$}} involved in the electron transfer, which we take to lie along the $z$ axis. The rotation angle $2\theta$ depends on the spin-orbit coupling $\vb*{\Lambda}_{\ce{AB}}=\Lambda_{\mathrm{AB}}\hat{\vb*{e}}_z$ and the spin-conserving electron transfer coupling $V_\mathrm{AB}$, \hl{between the SOMOs}
\begin{align}\label{eq-theta}
    \theta = \atan(\frac{\Lambda_{\mathrm{AB}}}{2V_\mathrm{AB}}).
\end{align}
Following from this, the electron transfer reaction is selective for a particular coherent superposition of the radical pair spin states $\ket{\psi_\theta}$, which is given by
\begin{align}
    \ket{\psi_\theta} = \cos\theta\ket{\mathrm{S}} + i \sin\theta \ket{\mathrm{T}_0},
\end{align}
where $\ket{\mathrm{S}}$ is the singlet spin-state and $\ket{\mathrm{T}_m}$ is triplet spin-state with projection quantum number $m=-1,0\text{ or }+1$.\cite{fay_chirality-induced_2021} Changing the chirality of the system changes the sign of $\vb*{\Lambda}_{\mathrm{AB}}$,\cite{dalum_theory_2019} and therefore changes the sign of the angle $\theta$. This means that the \ce{[A^$\bullet $B^$\bullet$]_{$R$}} system undergoes reactions selective for the $\ket{\psi_{+\theta}}$ spin state and the \ce{[A^$\bullet $B^$\bullet$]_{$S$}} intermediate undergoes reactions selective for the $\ket{\psi_{-\theta}}$ spin state.\cite{fay_chirality-induced_2021}
A more detailed discussion of the theory behind this chirality dependent spin selectivity can be found in Ref.~\onlinecite{fay_chirality-induced_2021}.

Now let us suppose that the radical pair \ce{[A^$\bullet $B^$\bullet$]} is initially formed spin-polarised state where the \ce{A^{$\bullet$}} radical is fully spin-polarised, and for simplicity we will take the spin polarisation to lie along the molecular $z$ axis (which is defined by the spin-orbit coupling vector $\vb*{\Lambda}_{\ce{AB}}$). The initial reduced density matrix for the system in either the \ce{[A^$\bullet $B^$\bullet$]_{$R$}} or \ce{[A^$\bullet $B^$\bullet$]_{$S$}} state is therefore
\begin{align}
    \hat{\sigma}_0 = \frac{1}{2} \left(\dyad{\uparrow_\mathrm{A}\uparrow_\mathrm{B}} +\dyad{\uparrow_\mathrm{A}\downarrow_\mathrm{B}} \right),
\end{align}
which can be re-written in terms of the coupled spin states as 
\begin{align}
    \hat{\sigma}_0 = \frac{1}{2}\left(\dyad{\mathrm{T}_{+1}} + \dyad{\Psi_0}\right)
\end{align}
where the $\ket{\Psi_0}$ state is
\begin{align}
    \ket{\Psi_0} = \frac{1}{\sqrt{2}}(\ket{\mathrm{S}} + \ket{\mathrm{T}_0}) = \ket{\uparrow_\mathrm{A}\downarrow_\mathrm{B}}.
\end{align}
The $\ket{\mathrm{T}_{+1}}$ fraction cannot undergo reactions to the \ce{{}^SP_{$R/S$}} products because $\braket{\psi_{\pm\theta}}{\mathrm{T}_+} = 0$, so this state will not be considered further. The $\ket{\Psi_0}$ fraction however does have a non-zero projection onto the reactive $\ket{{\psi_{\pm\theta}}}$ states
\begin{align}
    \braket{{\psi_{\pm\theta}}}{\Psi_0} = \frac{1}{\sqrt{2}}e^{\mp i\theta}.
\end{align}
The initial rate of reaction for the $R/S$ states is proportional to $|\braket{{\psi_{\pm\theta}}}{\Psi_0}|^2 = 1/2$, so initially the rates of formation of $R$ and $S$ enantiomers will be identical. However there will also exist an exchange interaction between the radicals \ce{A^{$\bullet$}} and \ce{B^{$\bullet$}}, which means the $\ket{\mathrm{S}}$ spin state will lie $+2J\hbar$ above the $\ket{\mathrm{T}_0}$ state.\cite{steiner_magnetic_1989} 
Therefore as the $\ket{\Psi_0}$ state evolves in time according to the Schr\"odinger equation a phase difference, $-(E_{\mathrm{S}}-E_{\mathrm{T}_0})t/\hbar$, manifests between the singlet and triplet components, and the time-evolved state $\ket{\Psi(t)}$ is
\begin{align}
    \ket{\Psi(t)} = \frac{1}{\sqrt{2}}\left(e^{-2i J t}\ket{\mathrm{S}} + \ket{\mathrm{T}_0}\right).
\end{align}
This time-evolved state now has different projections onto the reactive $\ket{\psi_{\pm\theta}}$ states,
\begin{align}
    \braket{{\psi_{\pm\theta}}}{\Psi (t)} = \frac{1}{\sqrt{2}} \left(e^{-2iJt}\cos(\theta) \mp i \sin(\theta)\right),
\end{align}
\hl{and the instantaneous reaction probability, which is proportional to the probability of finding the radical pair in the reactive state $\ket{\psi_{\pm\theta}}$, $|\braket{{\psi_{\pm\theta}}}{\Psi (t)}|^2$, is given by}
\begin{align}
    \left|\braket{{\psi_{\pm\theta}}}{\Psi (t)}\right|^2 = \frac{1}{2} \left(1 \pm \sin(2\theta)\sin(2J t)\right).
\end{align}
So we see that coherent evolution of the spin-polarised state due to the exchange interaction initially enhances the reaction probability for the $R$ ($+$) pathway over the $S$ ($-$) pathway, \hl{assuming $\theta > 0$}. If the $R$ and $S$ intermediates can also interconvert, \hl{as is often the case for high energy radical intermediates with partially filled orbitals,\cite{poh_photoredox_2025}} then this means the excess population in the \ce{[A^$\bullet $B^$\bullet$]_{$S$}} state after the initial reaction can replenish the \ce{[A^$\bullet $B^$\bullet$]_{$R$}} state, overall leading to an excess of the final \ce{{}^SP_{$R$}} product over the \ce{{}^SP_{$S$}} product. \hl{In summary the enantioselectivity of the spin-polarised radical recombination is controlled by the spin-orbit coupling between SOMOs on A and B, $\Lambda_{\mathrm{AB}}$, the direct coupling $V_{\mathrm{AB}}$, and the energy splitting between singlet and triplet radical pair states $2J$ and the coherent spin-dynamics of the radical pair that $2J$ induces, which creates an imbalance in instantaneous reaction rates.}

\hl{In the supporting information details are given of how to synthesise the ideas above into a quantitative model, which is illustrated graphically in Fig.~\ref{fig-scheme}(b). In this scheme, the $R$ and $S$ radical pair states are highlighted in red and blue. Each of these radical pairs is described with a spin density operator $\hat{\sigma}_{R/S}(t)$, which evolves coherently under a radical-pair spin Hamiltonian $\hat{H}_{+/-} \equiv \hat{H}_{R/S}$. These density operators can interconvert at a rate $k_\mathrm{f}$, the racemization rate of the radical pair, and the recombination to the singlet $R/S$ product state occurs at a rate $k_\mathrm{r}$ selectively from the $\ket{\psi_{\pm\theta}}$ states for the two enantiomers. This is described through chirality-dependent spin-selective reaction terms in the equation for $\dv{t}\hat{\sigma}_{R/S}(t)$ of the form $-(k_\mathrm{r}/2)\{\dyad{\psi_{+\theta}},\hat{\sigma}_{R}(t)\}$  for $\hat{\sigma}_{R}(t)$ and $-(k_\mathrm{r}/2)\{\dyad{\psi_{-\theta}},\hat{\sigma}_{S}(t)\}$  for $\hat{\sigma}_{S}(t)$, where $\{\hat{A},\hat{B}\}$ is the anticommutator. The possibility of spin-independent quenching of the radical pair intermediate at a rate $k_\mathrm{Q}$ is also considered later.} 

\begin{figure}
    \centering
    \includegraphics[width=0.8\linewidth]{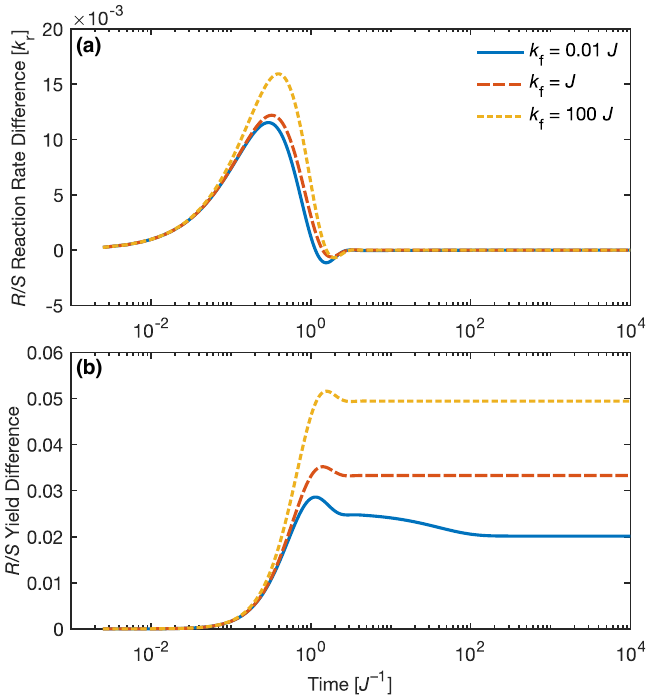}
   {\footnotesize \caption{Simulations of the time-dependent rate and yield differences for a radical pair initially in the $\hat{\sigma}_0 = \dyad{\uparrow_\mathrm{A} \downarrow_\mathrm{B}}$ state simulated using Eq.~S1. The model parameters are set to $k_\mathrm{r} = 4 J$, $\epsilon = 0$ and $\theta = 0.05$, and $k_\mathrm{f}$ is set to $100J$ (yellow short-dashed lines), $J$ (red long-dashed lines) and $0.01 J$ (blue solid lines). (a) Time-dependent reaction rate difference in units of $k_\mathrm{r}$ between $R$ and $S$ channels, i.e. $\Tr[\dyad{\psi_{+\theta}} \hat{\sigma}_R(t)]-\Tr[\dyad{\psi_{-\theta}} \hat{\sigma}_S(t)]$. (b) Time-dependent yield difference, $\phi_R(t)-\phi_S(t)$ \hl{(see supporting information for the definition of $\phi_{R/S}(t)$)}.}\label{fig-model-kf}}
    \vspace{0pt}
\end{figure}

In Fig.~\ref{fig-model-kf}(a) the instantaneous $R/S$ reaction probability difference, $\Tr[\dyad{\psi_{+\theta}} \hat{\sigma}_R(t)]-\Tr[\dyad{\psi_{-\theta}} \hat{\sigma}_S(t)]$, is shown for this model across a range of \hl{racemization rates} between $R$ and $S$ precursor states, $k_{\mathrm{f}}$, starting from the $\ket{\Psi_0}$ state (details of the quantitative modelling are given in the supporting information). \hl{In this example we set $\theta = 0.05$, which corresponds to $\Lambda_{\mathrm{AB}}/V_{\mathrm{AB}} \approx 10$, so assuming a typical spin-orbit coupling of $\Lambda_{\mathrm{AB}} \approx 1\ \text{cm}^{-1}$,\cite{poh_photoredox_2025} this corresponds to $V_{\mathrm{AB}} \approx 10\ \text{cm}^{-1}$, so this example is at the higher end of what is typical in organic radical pair systems, but still in a physically plausible regime.} This numerical calculation illustrates the above argument, showing a clear asymmetry in the $R$ and $S$ reaction rates \hl{for a small ratio of the spin-orbit coupling to direct electron transfer coupling, $\Lambda_{\mathrm{AB}}/V_{\mathrm{AB}}$ as is typical for organic radical pair systems}. Note that flipping the initial spin-polarisation corresponds to changing the sign of the $\ket{\mathrm{S}}$ component of the initial spin state, which is equivalent to an initial phase shift by $\pi$, which flips the selectively in the instantaneous reaction rate from $R$ to $S$.


Quantitative expressions for enantiomer yield asymmetry can be obtained for the full reaction scheme in Fig.~\ref{fig-scheme} \hl{through the stochastic Liouville equation formalism}, as is outlined in the supporting information. 
The only term appearing in these expressions below which we have not discussed above is $\epsilon$ measures an effective ``superexchange'' spin-orbit coupling between unpaired electrons that naturally appears \hl{emerges in the $R/S$ radical pair spin Hamiltonians $\hat{H}_{\pm} = 2\hbar J \dyad{\mathrm{S}} + \hbar \epsilon \dyad{\psi_{\pm\theta}}$ as a Lamb-shift type term.\cite{fay_chirality-induced_2021,fay_spin_2023} It will be demonstrated shortly that this term can enhance enantioselectivity even when spin-orbit coupling contributions to electron transfers are very small.}
\hl{Dipolar coupling between electron spins is ignored in this model for simplicity, and since for a close radical pair exchange coupling $J$ is expected to be much stronger than dipolar coupling.} We further assume that the initial radical \ce{A^$\bullet $} only has a spin-polarisation of $p_\mathrm{A}$, $1$ being fully spin-up and $-1$ being fully spin down, and the angle between this spin polarisation axis and the molecular spin-orbit coupling axis is $\chi$. After solving the Stochastic Liouville equations (with $k_\mathrm{Q} = 0$) an analytic expression for the yield difference between $R$ and $S$ reaction pathways is found, $\Delta\Phi = \Phi_R - \Phi_S$, where $\Phi_{R/S}$ is the final population on ${}^\mathrm{S}\ce{P}_{R/S}$ states in the scheme in Fig.~\ref{fig-scheme}. This is given by
\begin{align}\label{eq-dyield}
    \Delta \Phi = \left(\frac{p_\mathrm{A}\coschi}{2}\right) \frac{ k_{\mathrm{f}} k_{\mathrm{r}} \sin (2 \theta ) (J (4 k_{\mathrm{f}}+k_{\mathrm{r}})+2 k_{\mathrm{f}} \epsilon  \cos (2 \theta ))}{f(J,\epsilon,k_{\mathrm{f}} , k_{\mathrm{r}},\theta)}
\end{align}
where $f(J,\epsilon,k_{\mathrm{f}} , k_{\mathrm{r}},\theta)$ is a somewhat complicated function of the model parameters which is given in full in the supporting information. The absolute yields of the different products are given by $\Phi_{R} = 1/4 + \Delta \Phi/2$ and $\Phi_{S} = 1/4 - \Delta \Phi/2$, and the yield of unreacted ${\mathrm{T}_{\pm 1}}$ radical pairs is $\Phi_{\mathrm{T}_{\pm 1}} = 1/2$. We see that spin-orbit coupling mediated electron transfer is essential because if $\theta = 0$, corresponding to no spin-orbit mediated transfer, then $\Delta \Phi =0 $. Likewise we also see that the yield difference depends on the spin polarization projected onto the molecular $z$ axis $p_\mathrm{A}\coschi$ being non-zero as well as the existence of either $J$ or $\epsilon$ coupling between the radicals. So asymmetric radical pair electron transfers can only occur when spin-polarisation, spin-orbit coupling and coherent spin dynamics are present.

In the limit of large $k_\mathrm{f}$, i.e. very rapid \hl{racemization of} the $R$ and $S$ precursor radical pair states, the yield difference reduces to a simpler form,
\begin{align}\label{eq-fastex}
    \Delta \Phi = \left(\frac{p_\mathrm{A}\coschi}{2}\right)\frac{2 k_{\mathrm{r}} \sin (2 \theta ) (2 J+\epsilon  \cos(2 \theta ))}{k_{\mathrm{r}}^2 + 4(2 J+\epsilon  \cos(2 \theta ))^2} + \mathcal{O}(k_\mathrm{f}^{-1}).
\end{align}
This fast-exchange limit is likely physically fairly relevant, because the \ce{[A^$\bullet $B^$\bullet$]_{$R/S$}} states are expected to be high in energy and the interconversion barrier should be small. The yield difference in this limit is maximised, when $k_\mathrm{r} = 2 (2J + \epsilon \cos(2\theta))$, where it can be found to be
\begin{align}
    \Delta\Phi_{\mathrm{max}} = \frac{p_\mathrm{A}\coschi\sin(2\theta)}{4}.
\end{align}
Because only half of radical pairs can react at all, the enantiomeric excess of the reaction is given by $\mathrm{ee} = 2\Delta\Phi$. The maximum value this can take is $\mathrm{ee} = 50\%$, which occurs for a fully polarised radical $|p_\mathrm{A}|=1$, when that polarisation is parallel to the spin-orbit coupling axis, $|\coschi| = 1$, and when spin-orbit and spin-conserving electron tunneling make equal contributions to the electron transfer, i.e. when $|\sin(2\theta)|=1$. \hl{However for typical organic radical pairs where $\Lambda_{\mathrm{AB}} \sim 1\ \text{cm}^{-1}$ and $V_{\mathrm{AB}} \sim 10\!-\!100\ \text{cm}^{-1}$, $|\sin(2\theta)| \lesssim 0.05$, so for real systems in the fast racemization limit the enantiomeric excess would likely be small.}

In the opposite limit, where the rate of interconversion between $R$ and $S$ precursors is slow, i.e. $k_\mathrm{f}\to 0$, we can also find a simple expression for the yield difference as 
\begin{align}\label{eq-slowex}
    \Delta \Phi = \left(\frac{p_\mathrm{A}\coschi}{2}\right)\frac{k_\mathrm{f}}{J \sin(2\theta)} + \mathcal{O}(k_\mathrm{f}^2).
\end{align}
It can be seen that in this limit smaller values of $J$ and $\theta$ enhance the enantiomeric excess of the reaction. In Fig.~\ref{fig-model-kf}(b) the time-dependent $R/S$ yield difference is shown for a model radical pair calculated, for a range of values of $k_\mathrm{f}$, spanning the slow to fast interconversion limits, and a physically reasonable value of $\theta = 0.05$ for an organic radical pair (corresponding to roughly a 5\% contribution of spin-orbit coupling to the electron transfer). This simple example illustrates that a reasonable enantiomeric excess can be achieved across a wide range of values of $k_\mathrm{f}$, from the slow to fast limits, although the enantiomeric excess is enhanced in the rapid $R/S$ precursor \hl{racemization} limit. 

\begin{figure}
	 \centering
	\includegraphics[width=0.8\linewidth]{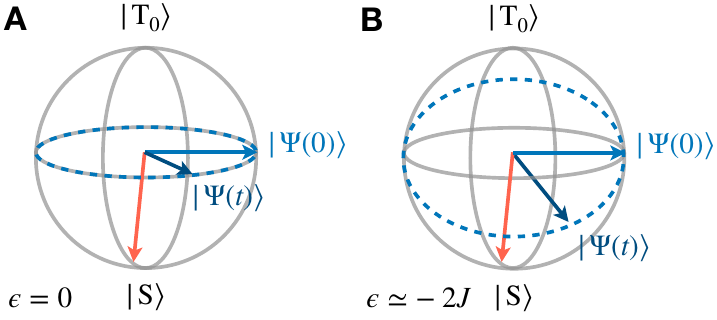}
	{\footnotesize\caption{Bloch sphere trajectories for the radical pair for different values of the spin superexchange coupling: a) $\epsilon = 0$ and b) $\epsilon ]simeq -2J$. The blue dashed line indicates the Bloch sphere trajectory and the blue arrows indicate the spin state of the radical pair at different times. The red arrow indicates the reactive state $\ket{\psi_\theta}$.}\label{fig-bloch}}
	\vspace{0pt}
\end{figure}

In the supporting information we also show how the inclusion of the spin-independent quenching process $k_\mathrm{Q}$ in Fig.~\ref{fig-scheme} leads to enantiomeric excesses. \hl{In this case we set $k_\mathrm{f} =0$ and $k_\mathrm{Q}>0$ in the scheme shown in Fig.~\ref{fig-scheme}.} \hl{In this case the chirality dependent spin-spin interactions, as described by the $\epsilon$ parameter, can play an important role in the radical pair spin dynamics. In the slow recombination limit, $k_\mathrm{r} \to 0$, most radical pairs are quenched, but for those that do react the enantiomeric excess in the products is
\begin{align}
	\mathrm{ee} = p_\mathrm{A}\coschi \frac{2 J k_{\mathrm{Q}} \sin (2 \theta )}{4 J^2+4 J \epsilon  \cos (2 \theta )+k_{\mathrm{Q}}^2+\epsilon ^2} + \mathcal{O}(k_\mathrm{r}).
\end{align}
This is maximised when $k_{\mathrm{Q}}^2 = 4 J^2+4 J \epsilon  \cos (2 \theta )+\epsilon ^2$ at $ p_\mathrm{A}\coschi (J\sin(2\theta)/k_\mathrm{Q})$, and because $\epsilon$ and $J$ can have different signs, this maximum value can exceed the $\sin(2\theta)/2$ limit above, and in fact it can be maximised for small $\theta$ when $\epsilon = -2 J $ giving a maximum enantiomeric excess of $p_\mathrm{A}\coschi/2$}. In principle for a fully polarised radical, the inclusion of quenching remarkably allows the enantiomeric excess to be maximised at 50\% independent of the magnitude of the spin-orbit coupling contribution to the electron transfer. \hl{So overall quenching may reduce the overall yield of the final product, but the presence of the chirality-dependent spin ``super-exchange'' type coupling $\epsilon$ can alter the coherent spin dynamics such that the enantiomeric excess is increased. This effects can be understood by considering the Bloch sphere where the poles correspond to the $\ket{\mathrm{T}_0}$ and $\ket{\mathrm{S}}$ states, as is illustrated in Fig.~\ref{fig-bloch}. The $x$ direction on the Bloch sphere corresponds to the spin-polarization and the $y$ direction corresponds to the imaginary part of the $\mathrm{S}$-$\mathrm{T}_0$ coherence, so the spin-polarized initial state lies along the $x$ direct and the reactive state lies in the $yz$ plane, close to $-z$. When $\epsilon =0$ the spin state follows a trajectory around the $xy$ plane, so the overlap with the reactive state can only change by an amount proportional to $\sin(2\theta)$. If instead $\epsilon \simeq -2J$ then the state now takes a different orbit around the Bloch sphere bringing it much closer to the reactive state at short times (or further away for the other enantiomer), thereby enabling much larger enantiomeric excesses to be achieved even for organic radical pairs with small spin-orbiot couplings.}

\hl{It is worth discussing the role of spin-relaxation in the enantioselective radical pair mechanism. Relaxation of the individual radical spins due to hyperfine interactions would reduce relax the spin-polarized initial state to the fully mixed unpolarized state, and likewise singlet-triplet dephasing due to fluctuations in exchange coupling with the inter-radical distance would destroy the coherence between singlet and triplet states, also inhibiting this effect.\cite{fay_electron_2019} If the radical pair spin-dynamics time-scale and lifetime are considerably longer than the spin-relaxation time, then spin-polarization would be too short-lived to give rise to enantioselective radical pair reactions. We therefore conclude that radical pairs separated by relatively short distances, which are therefore strongly coupled and recombine fast, would be more likely to show enantioselectivity. We also note that the theory in Ref.~\cite{fay_chirality-induced_2021} also predicts chirality-dependent spin-relaxation; this could lead to enhancement of enantioselectivity, through a mechanism analogous to noise-assisted transport in photosynthesis\cite{caruso_highly_2009}. However we reserve a more detailed analysis of spin-relaxation effects to a future publication. }

\begin{figure}[ht]
	\includegraphics[width=0.9\linewidth]{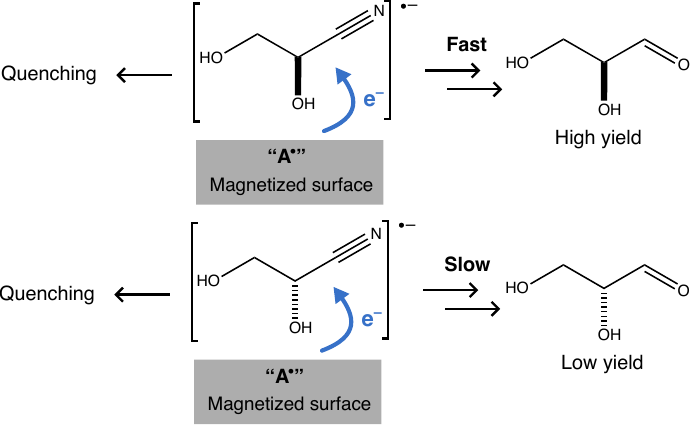}
	{\footnotesize\caption{An illustration of how a radical pair intermediate could lead chiral symmetry breaking in prebiotic reduction of glyceronitrile.}\label{fig-glyc}}
\end{figure}

Having established that electron transfer reactions of radical pairs can be in principle be enantioselective, it seems pertinent to now consider what could produce the required spin-polarisation and what reactions could show this spin-induced enantioselectivity. Firstly, turning to the question of generating spin-polarized radicals, a common source of these could be redox at magnetized ferromagnetic solid surfaces,\cite{ozturk_origin_2023,bloom_chiral_2024} where electrons transferred to or from molecules on the surface would have to be polarized in a particular direction due to the imbalance in the density of electronic states between up and down spins in the ferromagnet. In fact, unpaired electrons in the ferromagnet surface could also potentially act directly as the spin polarised \ce{A^{$\bullet$}} radical in the proposed scheme. \hl{The surface \ce{A^{$\bullet$}} radical could take the form of a secondary species on the surface which is reduced prior to reducing the \ce{B^{$\bullet$}} radical, or alternatively small polarons could potentially form on the surface of materials, particularly in response to absorption of a photon above the band gap of a material, which could act as the \ce{A^{$\bullet$}} radical.\cite{ren_recent_2024} These secondary surface species would have spins more isolated from the spins in the ferromagnet so could be less susceptible to spin relaxation. 
}

One potential example of the enantioselective radical-pair mechanism of particular relevance to the emergence of homochirality in Nature is the two-electron reduction of glyceronitrile, \hl{which could occur at the surface of magnetite when light is present to excite electrons from the magnetite surface}.\cite{ozturk_origins_2022,ritson_synthesis_2013} Assuming this two-electron reduction occurs sequentially, this\hl{ could} proceed via a radical pair type intermediate \hl{consisting of the partially reduced glyceronitrile radical anion (formed by an initial one electron reduction), and a spin polarized radical originating from the magnetized surface}. \hl{This is depicted in Fig.~\ref{fig-glyc}, where we assume that competition between quenching of the partially reduced species and spin-selective electron-transfer to the final reduction product produces enantioslectivity.} \hl{In this case the ``magnetized surface'' species that could act as the spin-polarized \ce{A^{$\bullet$}} radical could be a polaron at the magnetite surface,\cite{schrupp_high-energy_2005} or a second species at the surface that is reduced to give a spin-polarized radical that reduces the glyceronitrile radical anion. The final reduction has to be spin-selective because the reduction product is closed shell. The yield of the two enantiomers of the final glyceraldehyde product could be different due to the spin-orbit coupling effects described above. For this mechanism to end up producing an overall enantiomeric excess in the products, the quenched radical pair would ultimately need to be racemized before the radical pair is reformed, or completely decomposed into achiral products.}   

It is thought that the resulting glyceraldehyde can then react to form RNA precursors which preferentially crystallise in enantiomerically pure forms, which could lead to amplification of the initial chiral symmetry breaking.\cite{ozturk_central_2023} In the new mechanism proposed in this work transient coherent spin dynamics, together with spin-orbit coupling and electron-electron interactions produces the enantioselectivity for the same reaction. It is also possible that both the Ozturk and Sasselov mechanism and the radical pair mechanism operated together to produce homochirality in Nature. This initial symmetry breaking could then be amplified processes such as by crystallisation of chiral products.\cite{blackmond_origin_2019,sallembien_possible_2022} Ozturk \textit{et al.} have more recently suggested that photoelectrons from magnetite may in fact be produced in too low a yield for this to occur,\cite{ozturk_central_2023} but noting that experiments have shown spin-polarisation induced enantioselectivity at magnetised surfaces at room temperature in the absence of photo-excitation, the ejection of photoelectrons does not seem to be a requirement for spin-polarisation induced enantioselectivity.\cite{metzger_dynamic_2021} In the supporting information we propose some other example reactions where the enantioselective radical pair mechanism could play a role, and we postulate that the enantioseletive radical pair mechanism could play a role in at least of the experiments performed by Metzger \textit{et al.}.\cite{metzger_dynamic_2021}

In this letter a mechanism by which electron spin polarization can induce enantioselective reactions has been developed, together with a simple quantitative theory for the effect. Parameters appearing in the model, namely rate constants and electronic couplings can be measured experimentally or calculated using \textit{ab initio} methods,\cite{lawrence_path_2020,blumberger_recent_2015,malrieu_magnetic_2014} and the framework is straightforwardly extensible to include additional physical processes, such as electron spin relaxation, and to account for spin-selectivity in higher spin systems.\cite{fay_spin-selective_2018,fay_electron_2019} This enables us to find bounds on the maximum enantiomeric excess that can be produced by this mechanism and it also allows us to propose experimental tests of the mechanism. \hl{Interestingly, because this effect results from coherent spin dynamics, Lamb-shift terms appearing in the theory as spin-orbit coupling mediated super exchange terms can give rise to large effects even when spin-orbit coupling only makes a very small contribution to the electron transfer process.} This enantioselective radical pair mechanism could well play a role in proposed two-electron reduction processes involved in prebiotic cyanosulfidic chemistry,\cite{ozturk_origins_2022} as well as other radical pair redox reactions, and the theory presented here provides a route to quantitative investigation of spin-polarized electron induced enantioselective reactions. \hl{Recent work in a similar vein has proposed that magnetic fields, can also give rise to enantioselective reactions in radical pair intermediates through a very similar mechanism to that proposed here.\cite{poh_photoredox_2025} This suggests that accounting for magnetic fields, as will be present near magnetized surfaces, could potentially enhance this mechanism further. While currently there is only very limited direct experimental evidence of spin-polarization leading to enantioselectivity, the author hopes that this theory provides guidance and motivation to devise experiments probing the role of spin-polarized electrons in enantioselective reactions. }

\vspace{-10pt}
\subsection*{Acknowledgements}
\vspace{-10pt}
\noindent TPF would like to thank Joseph Lawrence, Alistair Sterling, Tomoyasu Mani, and Johan Runeson for their comments on this manuscript. 
\vspace{-10pt}
\subsection*{Data Availability}
\vspace{-10pt}
\noindent All data relevant to this work are available within the text and supporting information.
\vspace{-10pt}
\subsection*{Conflicts of interest}
\noindent The author has no conflict of interest to declare.
\vspace{-10pt}
\subsection*{Supporting Information}
\vspace{-10pt}
\noindent In the supporting information further details of the Stochastic Liouville equation model for the enantioselective radical pair mechanism are discussed, as well as details of the analytic results for quantum yields. Additionally some other specific classes of chemical reaction involving radical pairs are proposed which could display spin-polarization dependent enantioselectivity, and a discussion of how to experimentally verify this mechanism.

\bibliography{refs}

\begin{thebibliography}{47}%
\makeatletter
\providecommand \@ifxundefined [1]{%
 \@ifx{#1\undefined}
}%
\providecommand \@ifnum [1]{%
 \ifnum #1\expandafter \@firstoftwo
 \else \expandafter \@secondoftwo
 \fi
}%
\providecommand \@ifx [1]{%
 \ifx #1\expandafter \@firstoftwo
 \else \expandafter \@secondoftwo
 \fi
}%
\providecommand \natexlab [1]{#1}%
\providecommand \enquote  [1]{``#1''}%
\providecommand \bibnamefont  [1]{#1}%
\providecommand \bibfnamefont [1]{#1}%
\providecommand \citenamefont [1]{#1}%
\providecommand \href@noop [0]{\@secondoftwo}%
\providecommand \href [0]{\begingroup \@sanitize@url \@href}%
\providecommand \@href[1]{\@@startlink{#1}\@@href}%
\providecommand \@@href[1]{\endgroup#1\@@endlink}%
\providecommand \@sanitize@url [0]{\catcode `\\12\catcode `\$12\catcode
  `\&12\catcode `\#12\catcode `\^12\catcode `\_12\catcode `\%12\relax}%
\providecommand \@@startlink[1]{}%
\providecommand \@@endlink[0]{}%
\providecommand \url  [0]{\begingroup\@sanitize@url \@url }%
\providecommand \@url [1]{\endgroup\@href {#1}{\urlprefix }}%
\providecommand \urlprefix  [0]{URL }%
\providecommand \Eprint [0]{\href }%
\providecommand \doibase [0]{https://doi.org/}%
\providecommand \selectlanguage [0]{\@gobble}%
\providecommand \bibinfo  [0]{\@secondoftwo}%
\providecommand \bibfield  [0]{\@secondoftwo}%
\providecommand \translation [1]{[#1]}%
\providecommand \BibitemOpen [0]{}%
\providecommand \bibitemStop [0]{}%
\providecommand \bibitemNoStop [0]{.\EOS\space}%
\providecommand \EOS [0]{\spacefactor3000\relax}%
\providecommand \BibitemShut  [1]{\csname bibitem#1\endcsname}%
\let\auto@bib@innerbib\@empty
\bibitem [{\citenamefont {Blackmond}(2019)}]{blackmond_origin_2019}%
  \BibitemOpen
  \bibfield  {author} {\bibinfo {author} {\bibfnamefont {D.~G.}\ \bibnamefont
  {Blackmond}},\ }\bibfield  {title} {\bibinfo {title} {The {Origin} of
  {Biological} {Homochirality}},\ }\href
  {https://doi.org/10.1101/cshperspect.a032540} {\bibfield  {journal} {\bibinfo
   {journal} {Cold Spring Harbor Perspectives in Biology}\ }\textbf {\bibinfo
  {volume} {11}},\ \bibinfo {pages} {a032540} (\bibinfo {year}
  {2019})}\BibitemShut {NoStop}%
\bibitem [{\citenamefont {Sallembien}\ \emph {et~al.}(2022)\citenamefont
  {Sallembien}, \citenamefont {Bouteiller}, \citenamefont {Crassous},\ and\
  \citenamefont {Raynal}}]{sallembien_possible_2022}%
  \BibitemOpen
  \bibfield  {author} {\bibinfo {author} {\bibfnamefont {Q.}~\bibnamefont
  {Sallembien}}, \bibinfo {author} {\bibfnamefont {L.}~\bibnamefont
  {Bouteiller}}, \bibinfo {author} {\bibfnamefont {J.}~\bibnamefont
  {Crassous}},\ and\ \bibinfo {author} {\bibfnamefont {M.}~\bibnamefont
  {Raynal}},\ }\bibfield  {title} {\bibinfo {title} {Possible chemical and
  physical scenarios towards biological homochirality},\ }\href
  {https://doi.org/10.1039/D1CS01179K} {\bibfield  {journal} {\bibinfo
  {journal} {Chemical Society Reviews}\ }\textbf {\bibinfo {volume} {51}},\
  \bibinfo {pages} {3436} (\bibinfo {year} {2022})}\BibitemShut {NoStop}%
\bibitem [{\citenamefont {Rau}(1983)}]{rau_asymmetric_1983}%
  \BibitemOpen
  \bibfield  {author} {\bibinfo {author} {\bibfnamefont {H.}~\bibnamefont
  {Rau}},\ }\bibfield  {title} {\bibinfo {title} {Asymmetric photochemistry in
  solution},\ }\href {https://doi.org/10.1021/cr00057a003} {\bibfield
  {journal} {\bibinfo  {journal} {Chemical Reviews}\ }\textbf {\bibinfo
  {volume} {83}},\ \bibinfo {pages} {535} (\bibinfo {year} {1983})}\BibitemShut
  {NoStop}%
\bibitem [{\citenamefont {Rikken}\ and\ \citenamefont
  {Raupach}(2000)}]{rikken_enantioselective_2000}%
  \BibitemOpen
  \bibfield  {author} {\bibinfo {author} {\bibfnamefont {G.~L. J.~A.}\
  \bibnamefont {Rikken}}\ and\ \bibinfo {author} {\bibfnamefont
  {E.}~\bibnamefont {Raupach}},\ }\bibfield  {title} {\bibinfo {title}
  {Enantioselective magnetochiral photochemistry},\ }\href
  {https://doi.org/10.1038/35016043} {\bibfield  {journal} {\bibinfo  {journal}
  {Nature}\ }\textbf {\bibinfo {volume} {405}},\ \bibinfo {pages} {932}
  (\bibinfo {year} {2000})}\BibitemShut {NoStop}%
\bibitem [{\citenamefont {Ozturk}\ and\ \citenamefont
  {Sasselov}(2022)}]{ozturk_origins_2022}%
  \BibitemOpen
  \bibfield  {author} {\bibinfo {author} {\bibfnamefont {S.~F.}\ \bibnamefont
  {Ozturk}}\ and\ \bibinfo {author} {\bibfnamefont {D.~D.}\ \bibnamefont
  {Sasselov}},\ }\bibfield  {title} {\bibinfo {title} {On the origins of
  life’s homochirality: {Inducing} enantiomeric excess with spin-polarized
  electrons},\ }\href {https://doi.org/10.1073/pnas.2204765119} {\bibfield
  {journal} {\bibinfo  {journal} {Proceedings of the National Academy of
  Sciences}\ }\textbf {\bibinfo {volume} {119}},\ \bibinfo {pages}
  {e2204765119} (\bibinfo {year} {2022})}\BibitemShut {NoStop}%
\bibitem [{\citenamefont {Ozturk}\ \emph
  {et~al.}(2023{\natexlab{a}})\citenamefont {Ozturk}, \citenamefont {Liu},
  \citenamefont {Sutherland},\ and\ \citenamefont
  {Sasselov}}]{ozturk_origin_2023}%
  \BibitemOpen
  \bibfield  {author} {\bibinfo {author} {\bibfnamefont {S.~F.}\ \bibnamefont
  {Ozturk}}, \bibinfo {author} {\bibfnamefont {Z.}~\bibnamefont {Liu}},
  \bibinfo {author} {\bibfnamefont {J.~D.}\ \bibnamefont {Sutherland}},\ and\
  \bibinfo {author} {\bibfnamefont {D.~D.}\ \bibnamefont {Sasselov}},\
  }\bibfield  {title} {\bibinfo {title} {Origin of biological homochirality by
  crystallization of an {RNA} precursor on a magnetic surface},\ }\href
  {https://doi.org/10.1126/sciadv.adg8274} {\bibfield  {journal} {\bibinfo
  {journal} {Science Advances}\ }\textbf {\bibinfo {volume} {9}},\ \bibinfo
  {pages} {eadg8274} (\bibinfo {year} {2023}{\natexlab{a}})}\BibitemShut
  {NoStop}%
\bibitem [{\citenamefont {Ozturk}\ \emph
  {et~al.}(2023{\natexlab{b}})\citenamefont {Ozturk}, \citenamefont
  {Sasselov},\ and\ \citenamefont {Sutherland}}]{ozturk_central_2023}%
  \BibitemOpen
  \bibfield  {author} {\bibinfo {author} {\bibfnamefont {S.~F.}\ \bibnamefont
  {Ozturk}}, \bibinfo {author} {\bibfnamefont {D.~D.}\ \bibnamefont
  {Sasselov}},\ and\ \bibinfo {author} {\bibfnamefont {J.~D.}\ \bibnamefont
  {Sutherland}},\ }\bibfield  {title} {\bibinfo {title} {The central dogma of
  biological homochirality: {How} does chiral information propagate in a
  prebiotic network?},\ }\href {https://doi.org/10.1063/5.0156527} {\bibfield
  {journal} {\bibinfo  {journal} {The Journal of Chemical Physics}\ }\textbf
  {\bibinfo {volume} {159}},\ \bibinfo {pages} {061102} (\bibinfo {year}
  {2023}{\natexlab{b}})}\BibitemShut {NoStop}%
\bibitem [{\citenamefont {Metzger}\ \emph {et~al.}(2020)\citenamefont
  {Metzger}, \citenamefont {Mishra}, \citenamefont {Bloom}, \citenamefont
  {Goren}, \citenamefont {Neubauer}, \citenamefont {Shmul}, \citenamefont
  {Wei}, \citenamefont {Yochelis}, \citenamefont {Tassinari}, \citenamefont
  {Fontanesi}, \citenamefont {Waldeck}, \citenamefont {Paltiel},\ and\
  \citenamefont {Naaman}}]{metzger_electron_2020}%
  \BibitemOpen
  \bibfield  {author} {\bibinfo {author} {\bibfnamefont {T.~S.}\ \bibnamefont
  {Metzger}}, \bibinfo {author} {\bibfnamefont {S.}~\bibnamefont {Mishra}},
  \bibinfo {author} {\bibfnamefont {B.~P.}\ \bibnamefont {Bloom}}, \bibinfo
  {author} {\bibfnamefont {N.}~\bibnamefont {Goren}}, \bibinfo {author}
  {\bibfnamefont {A.}~\bibnamefont {Neubauer}}, \bibinfo {author}
  {\bibfnamefont {G.}~\bibnamefont {Shmul}}, \bibinfo {author} {\bibfnamefont
  {J.}~\bibnamefont {Wei}}, \bibinfo {author} {\bibfnamefont {S.}~\bibnamefont
  {Yochelis}}, \bibinfo {author} {\bibfnamefont {F.}~\bibnamefont {Tassinari}},
  \bibinfo {author} {\bibfnamefont {C.}~\bibnamefont {Fontanesi}}, \bibinfo
  {author} {\bibfnamefont {D.~H.}\ \bibnamefont {Waldeck}}, \bibinfo {author}
  {\bibfnamefont {Y.}~\bibnamefont {Paltiel}},\ and\ \bibinfo {author}
  {\bibfnamefont {R.}~\bibnamefont {Naaman}},\ }\bibfield  {title} {\bibinfo
  {title} {The {Electron} {Spin} as a {Chiral} {Reagent}},\ }\href
  {https://doi.org/10.1002/anie.201911400} {\bibfield  {journal} {\bibinfo
  {journal} {Angewandte Chemie International Edition}\ }\textbf {\bibinfo
  {volume} {59}},\ \bibinfo {pages} {1653} (\bibinfo {year}
  {2020})}\BibitemShut {NoStop}%
\bibitem [{\citenamefont {Metzger}\ \emph {et~al.}(2021)\citenamefont
  {Metzger}, \citenamefont {Siam}, \citenamefont {Kolodny}, \citenamefont
  {Goren}, \citenamefont {Sukenik}, \citenamefont {Yochelis}, \citenamefont
  {Abu-Reziq}, \citenamefont {Avnir},\ and\ \citenamefont
  {Paltiel}}]{metzger_dynamic_2021}%
  \BibitemOpen
  \bibfield  {author} {\bibinfo {author} {\bibfnamefont {T.~S.}\ \bibnamefont
  {Metzger}}, \bibinfo {author} {\bibfnamefont {R.}~\bibnamefont {Siam}},
  \bibinfo {author} {\bibfnamefont {Y.}~\bibnamefont {Kolodny}}, \bibinfo
  {author} {\bibfnamefont {N.}~\bibnamefont {Goren}}, \bibinfo {author}
  {\bibfnamefont {N.}~\bibnamefont {Sukenik}}, \bibinfo {author} {\bibfnamefont
  {S.}~\bibnamefont {Yochelis}}, \bibinfo {author} {\bibfnamefont
  {R.}~\bibnamefont {Abu-Reziq}}, \bibinfo {author} {\bibfnamefont
  {D.}~\bibnamefont {Avnir}},\ and\ \bibinfo {author} {\bibfnamefont
  {Y.}~\bibnamefont {Paltiel}},\ }\bibfield  {title} {\bibinfo {title} {Dynamic
  {Spin}-{Controlled} {Enantioselective} {Catalytic} {Chiral} {Reactions}},\
  }\href {https://doi.org/10.1021/acs.jpclett.1c01518} {\bibfield  {journal}
  {\bibinfo  {journal} {The Journal of Physical Chemistry Letters}\ }\textbf
  {\bibinfo {volume} {12}},\ \bibinfo {pages} {5469} (\bibinfo {year}
  {2021})}\BibitemShut {NoStop}%
\bibitem [{\citenamefont {Bhowmick}\ \emph {et~al.}(2022)\citenamefont
  {Bhowmick}, \citenamefont {Das}, \citenamefont {Santra}, \citenamefont
  {Mondal}, \citenamefont {Tassinari}, \citenamefont {Schwarz}, \citenamefont
  {Diesendruck},\ and\ \citenamefont {Naaman}}]{bhowmick_spin-induced_2022}%
  \BibitemOpen
  \bibfield  {author} {\bibinfo {author} {\bibfnamefont {D.~K.}\ \bibnamefont
  {Bhowmick}}, \bibinfo {author} {\bibfnamefont {T.~K.}\ \bibnamefont {Das}},
  \bibinfo {author} {\bibfnamefont {K.}~\bibnamefont {Santra}}, \bibinfo
  {author} {\bibfnamefont {A.~K.}\ \bibnamefont {Mondal}}, \bibinfo {author}
  {\bibfnamefont {F.}~\bibnamefont {Tassinari}}, \bibinfo {author}
  {\bibfnamefont {R.}~\bibnamefont {Schwarz}}, \bibinfo {author} {\bibfnamefont
  {C.~E.}\ \bibnamefont {Diesendruck}},\ and\ \bibinfo {author} {\bibfnamefont
  {R.}~\bibnamefont {Naaman}},\ }\bibfield  {title} {\bibinfo {title}
  {Spin-induced asymmetry reaction—{The} formation of asymmetric carbon by
  electropolymerization},\ }\href {https://doi.org/10.1126/sciadv.abq2727}
  {\bibfield  {journal} {\bibinfo  {journal} {Science Advances}\ }\textbf
  {\bibinfo {volume} {8}},\ \bibinfo {pages} {eabq2727} (\bibinfo {year}
  {2022})}\BibitemShut {NoStop}%
\bibitem [{\citenamefont {Bloom}\ \emph {et~al.}(2024)\citenamefont {Bloom},
  \citenamefont {Paltiel}, \citenamefont {Naaman},\ and\ \citenamefont
  {Waldeck}}]{bloom_chiral_2024}%
  \BibitemOpen
  \bibfield  {author} {\bibinfo {author} {\bibfnamefont {B.~P.}\ \bibnamefont
  {Bloom}}, \bibinfo {author} {\bibfnamefont {Y.}~\bibnamefont {Paltiel}},
  \bibinfo {author} {\bibfnamefont {R.}~\bibnamefont {Naaman}},\ and\ \bibinfo
  {author} {\bibfnamefont {D.~H.}\ \bibnamefont {Waldeck}},\ }\bibfield
  {title} {\bibinfo {title} {Chiral {Induced} {Spin} {Selectivity}},\ }\href
  {https://doi.org/10.1021/acs.chemrev.3c00661} {\bibfield  {journal} {\bibinfo
   {journal} {Chemical Reviews}\ }\textbf {\bibinfo {volume} {124}},\ \bibinfo
  {pages} {1950} (\bibinfo {year} {2024})}\BibitemShut {NoStop}%
\bibitem [{\citenamefont {Wu}\ and\ \citenamefont
  {Subotnik}(2021)}]{wu_electronic_2021}%
  \BibitemOpen
  \bibfield  {author} {\bibinfo {author} {\bibfnamefont {Y.}~\bibnamefont
  {Wu}}\ and\ \bibinfo {author} {\bibfnamefont {J.~E.}\ \bibnamefont
  {Subotnik}},\ }\bibfield  {title} {\bibinfo {title} {Electronic spin
  separation induced by nuclear motion near conical intersections},\ }\href
  {https://doi.org/10.1038/s41467-020-20831-8} {\bibfield  {journal} {\bibinfo
  {journal} {Nature Communications}\ }\textbf {\bibinfo {volume} {12}},\
  \bibinfo {pages} {700} (\bibinfo {year} {2021})}\BibitemShut {NoStop}%
\bibitem [{\citenamefont {Bian}\ and\ \citenamefont
  {Subotnik}(2024)}]{bian_spin-dependent_2024}%
  \BibitemOpen
  \bibfield  {author} {\bibinfo {author} {\bibfnamefont {X.}~\bibnamefont
  {Bian}}\ and\ \bibinfo {author} {\bibfnamefont {J.~E.}\ \bibnamefont
  {Subotnik}},\ }\bibfield  {title} {\bibinfo {title} {Spin-{Dependent}
  {Stereochemistry}: {A} {Nonadiabatic} {Quantum} {Dynamics} {Case} {Study} of
  {S} + {H} $_{\textrm{2}}$ → {SH} + {H} {Reaction}},\ }\href
  {https://doi.org/10.1021/acs.jpclett.3c03344} {\bibfield  {journal} {\bibinfo
   {journal} {The Journal of Physical Chemistry Letters}\ }\textbf {\bibinfo
  {volume} {15}},\ \bibinfo {pages} {3434} (\bibinfo {year}
  {2024})}\BibitemShut {NoStop}%
\bibitem [{\citenamefont {Ozturk}\ \emph
  {et~al.}(2023{\natexlab{c}})\citenamefont {Ozturk}, \citenamefont {Bhowmick},
  \citenamefont {Kapon}, \citenamefont {Sang}, \citenamefont {Kumar},
  \citenamefont {Paltiel}, \citenamefont {Naaman},\ and\ \citenamefont
  {Sasselov}}]{ozturk_chirality-induced_2023}%
  \BibitemOpen
  \bibfield  {author} {\bibinfo {author} {\bibfnamefont {S.~F.}\ \bibnamefont
  {Ozturk}}, \bibinfo {author} {\bibfnamefont {D.~K.}\ \bibnamefont
  {Bhowmick}}, \bibinfo {author} {\bibfnamefont {Y.}~\bibnamefont {Kapon}},
  \bibinfo {author} {\bibfnamefont {Y.}~\bibnamefont {Sang}}, \bibinfo {author}
  {\bibfnamefont {A.}~\bibnamefont {Kumar}}, \bibinfo {author} {\bibfnamefont
  {Y.}~\bibnamefont {Paltiel}}, \bibinfo {author} {\bibfnamefont
  {R.}~\bibnamefont {Naaman}},\ and\ \bibinfo {author} {\bibfnamefont {D.~D.}\
  \bibnamefont {Sasselov}},\ }\bibfield  {title} {\bibinfo {title}
  {Chirality-induced avalanche magnetization of magnetite by an {RNA}
  precursor},\ }\href {https://doi.org/10.1038/s41467-023-42130-8} {\bibfield
  {journal} {\bibinfo  {journal} {Nature Communications}\ }\textbf {\bibinfo
  {volume} {14}},\ \bibinfo {pages} {6351} (\bibinfo {year}
  {2023}{\natexlab{c}})},\ \bibinfo {note} {publisher: Springer Science and
  Business Media LLC}\BibitemShut {NoStop}%
\bibitem [{\citenamefont {Steiner}\ and\ \citenamefont
  {Ulrich}(1989)}]{steiner_magnetic_1989}%
  \BibitemOpen
  \bibfield  {author} {\bibinfo {author} {\bibfnamefont {U.~E.}\ \bibnamefont
  {Steiner}}\ and\ \bibinfo {author} {\bibfnamefont {T.}~\bibnamefont
  {Ulrich}},\ }\bibfield  {title} {\bibinfo {title} {Magnetic field effects in
  chemical kinetics and related phenomena},\ }\href
  {https://doi.org/10.1021/cr00091a003} {\bibfield  {journal} {\bibinfo
  {journal} {Chemical Reviews}\ }\textbf {\bibinfo {volume} {89}},\ \bibinfo
  {pages} {51} (\bibinfo {year} {1989})}\BibitemShut {NoStop}%
\bibitem [{\citenamefont {Mani}(2022)}]{mani_molecular_2022}%
  \BibitemOpen
  \bibfield  {author} {\bibinfo {author} {\bibfnamefont {T.}~\bibnamefont
  {Mani}},\ }\bibfield  {title} {\bibinfo {title} {Molecular qubits based on
  photogenerated spin-correlated radical pairs for quantum sensing},\ }\href
  {https://doi.org/10.1063/5.0084072} {\bibfield  {journal} {\bibinfo
  {journal} {Chemical Physics Reviews}\ }\textbf {\bibinfo {volume} {3}},\
  \bibinfo {pages} {021301} (\bibinfo {year} {2022})}\BibitemShut {NoStop}%
\bibitem [{\citenamefont {Maeda}\ \emph {et~al.}(2008)\citenamefont {Maeda},
  \citenamefont {Henbest}, \citenamefont {Cintolesi}, \citenamefont {Kuprov},
  \citenamefont {Rodgers}, \citenamefont {Liddell}, \citenamefont {Gust},
  \citenamefont {Timmel},\ and\ \citenamefont {Hore}}]{maeda_chemical_2008}%
  \BibitemOpen
  \bibfield  {author} {\bibinfo {author} {\bibfnamefont {K.}~\bibnamefont
  {Maeda}}, \bibinfo {author} {\bibfnamefont {K.~B.}\ \bibnamefont {Henbest}},
  \bibinfo {author} {\bibfnamefont {F.}~\bibnamefont {Cintolesi}}, \bibinfo
  {author} {\bibfnamefont {I.}~\bibnamefont {Kuprov}}, \bibinfo {author}
  {\bibfnamefont {C.~T.}\ \bibnamefont {Rodgers}}, \bibinfo {author}
  {\bibfnamefont {P.~A.}\ \bibnamefont {Liddell}}, \bibinfo {author}
  {\bibfnamefont {D.}~\bibnamefont {Gust}}, \bibinfo {author} {\bibfnamefont
  {C.~R.}\ \bibnamefont {Timmel}},\ and\ \bibinfo {author} {\bibfnamefont
  {P.~J.}\ \bibnamefont {Hore}},\ }\bibfield  {title} {\bibinfo {title}
  {Chemical compass model of avian magnetoreception},\ }\href
  {https://doi.org/10.1038/nature06834} {\bibfield  {journal} {\bibinfo
  {journal} {Nature}\ }\textbf {\bibinfo {volume} {453}},\ \bibinfo {pages}
  {387} (\bibinfo {year} {2008})}\BibitemShut {NoStop}%
\bibitem [{\citenamefont {Hore}\ and\ \citenamefont
  {Mouritsen}(2016)}]{hore_radical-pair_2016}%
  \BibitemOpen
  \bibfield  {author} {\bibinfo {author} {\bibfnamefont {P.~J.}\ \bibnamefont
  {Hore}}\ and\ \bibinfo {author} {\bibfnamefont {H.}~\bibnamefont
  {Mouritsen}},\ }\bibfield  {title} {\bibinfo {title} {The {Radical}-{Pair}
  {Mechanism} of {Magnetoreception}},\ }\href
  {https://doi.org/10.1146/annurev-biophys-032116-094545} {\bibfield  {journal}
  {\bibinfo  {journal} {Annual Review of Biophysics}\ }\textbf {\bibinfo
  {volume} {45}},\ \bibinfo {pages} {299} (\bibinfo {year} {2016})}\BibitemShut
  {NoStop}%
\bibitem [{\citenamefont {Carmeli}\ \emph {et~al.}(2014)\citenamefont
  {Carmeli}, \citenamefont {Kumar}, \citenamefont {Heifler}, \citenamefont
  {Carmeli},\ and\ \citenamefont {Naaman}}]{carmeli_spin_2014}%
  \BibitemOpen
  \bibfield  {author} {\bibinfo {author} {\bibfnamefont {I.}~\bibnamefont
  {Carmeli}}, \bibinfo {author} {\bibfnamefont {K.~S.}\ \bibnamefont {Kumar}},
  \bibinfo {author} {\bibfnamefont {O.}~\bibnamefont {Heifler}}, \bibinfo
  {author} {\bibfnamefont {C.}~\bibnamefont {Carmeli}},\ and\ \bibinfo {author}
  {\bibfnamefont {R.}~\bibnamefont {Naaman}},\ }\bibfield  {title} {\bibinfo
  {title} {Spin {Selectivity} in {Electron} {Transfer} in {Photosystem} {I}},\
  }\href {https://doi.org/10.1002/anie.201404382} {\bibfield  {journal}
  {\bibinfo  {journal} {Angewandte Chemie International Edition}\ }\textbf
  {\bibinfo {volume} {53}},\ \bibinfo {pages} {8953} (\bibinfo {year}
  {2014})}\BibitemShut {NoStop}%
\bibitem [{\citenamefont {Eckvahl}\ \emph {et~al.}(2023)\citenamefont
  {Eckvahl}, \citenamefont {Tcyrulnikov}, \citenamefont {Chiesa}, \citenamefont
  {Bradley}, \citenamefont {Young}, \citenamefont {Carretta}, \citenamefont
  {Krzyaniak},\ and\ \citenamefont {Wasielewski}}]{eckvahl_direct_2023}%
  \BibitemOpen
  \bibfield  {author} {\bibinfo {author} {\bibfnamefont {H.~J.}\ \bibnamefont
  {Eckvahl}}, \bibinfo {author} {\bibfnamefont {N.~A.}\ \bibnamefont
  {Tcyrulnikov}}, \bibinfo {author} {\bibfnamefont {A.}~\bibnamefont {Chiesa}},
  \bibinfo {author} {\bibfnamefont {J.~M.}\ \bibnamefont {Bradley}}, \bibinfo
  {author} {\bibfnamefont {R.~M.}\ \bibnamefont {Young}}, \bibinfo {author}
  {\bibfnamefont {S.}~\bibnamefont {Carretta}}, \bibinfo {author}
  {\bibfnamefont {M.~D.}\ \bibnamefont {Krzyaniak}},\ and\ \bibinfo {author}
  {\bibfnamefont {M.~R.}\ \bibnamefont {Wasielewski}},\ }\bibfield  {title}
  {\bibinfo {title} {Direct observation of chirality-induced spin selectivity
  in electron donor–acceptor molecules},\ }\href
  {https://doi.org/10.1126/science.adj5328} {\bibfield  {journal} {\bibinfo
  {journal} {Science}\ }\textbf {\bibinfo {volume} {382}},\ \bibinfo {pages}
  {197} (\bibinfo {year} {2023})}\BibitemShut {NoStop}%
\bibitem [{\citenamefont {Eckvahl}\ \emph {et~al.}(2024)\citenamefont
  {Eckvahl}, \citenamefont {Copley}, \citenamefont {Young}, \citenamefont
  {Krzyaniak},\ and\ \citenamefont {Wasielewski}}]{eckvahl_detecting_2024}%
  \BibitemOpen
  \bibfield  {author} {\bibinfo {author} {\bibfnamefont {H.~J.}\ \bibnamefont
  {Eckvahl}}, \bibinfo {author} {\bibfnamefont {G.}~\bibnamefont {Copley}},
  \bibinfo {author} {\bibfnamefont {R.~M.}\ \bibnamefont {Young}}, \bibinfo
  {author} {\bibfnamefont {M.~D.}\ \bibnamefont {Krzyaniak}},\ and\ \bibinfo
  {author} {\bibfnamefont {M.~R.}\ \bibnamefont {Wasielewski}},\ }\bibfield
  {title} {\bibinfo {title} {Detecting {Chirality}-{Induced} {Spin}
  {Selectivity} in {Randomly} {Oriented} {Radical} {Pairs} {Photogenerated} by
  {Hole} {Transfer}},\ }\href {https://doi.org/10.1021/jacs.4c08706} {\bibfield
   {journal} {\bibinfo  {journal} {Journal of the American Chemical Society}\
  }\textbf {\bibinfo {volume} {146}},\ \bibinfo {pages} {24125} (\bibinfo
  {year} {2024})}\BibitemShut {NoStop}%
\bibitem [{\citenamefont {Fay}(2021)}]{fay_chirality-induced_2021}%
  \BibitemOpen
  \bibfield  {author} {\bibinfo {author} {\bibfnamefont {T.~P.}\ \bibnamefont
  {Fay}},\ }\bibfield  {title} {\bibinfo {title} {Chirality-{Induced} {Spin}
  {Coherence} in {Electron} {Transfer} {Reactions}},\ }\href
  {https://doi.org/10.1021/acs.jpclett.1c00009} {\bibfield  {journal} {\bibinfo
   {journal} {The Journal of Physical Chemistry Letters}\ }\textbf {\bibinfo
  {volume} {12}},\ \bibinfo {pages} {1407} (\bibinfo {year}
  {2021})}\BibitemShut {NoStop}%
\bibitem [{\citenamefont {Fay}\ and\ \citenamefont
  {Limmer}(2021)}]{fay_origin_2021}%
  \BibitemOpen
  \bibfield  {author} {\bibinfo {author} {\bibfnamefont {T.~P.}\ \bibnamefont
  {Fay}}\ and\ \bibinfo {author} {\bibfnamefont {D.~T.}\ \bibnamefont
  {Limmer}},\ }\bibfield  {title} {\bibinfo {title} {Origin of {Chirality}
  {Induced} {Spin} {Selectivity} in {Photoinduced} {Electron} {Transfer}},\
  }\href {https://doi.org/10.1021/acs.nanolett.1c02370} {\bibfield  {journal}
  {\bibinfo  {journal} {Nano Letters}\ }\textbf {\bibinfo {volume} {21}},\
  \bibinfo {pages} {6696} (\bibinfo {year} {2021})}\BibitemShut {NoStop}%
\bibitem [{\citenamefont {Fay}\ and\ \citenamefont
  {Limmer}(2023)}]{fay_spin_2023}%
  \BibitemOpen
  \bibfield  {author} {\bibinfo {author} {\bibfnamefont {T.~P.}\ \bibnamefont
  {Fay}}\ and\ \bibinfo {author} {\bibfnamefont {D.~T.}\ \bibnamefont
  {Limmer}},\ }\bibfield  {title} {\bibinfo {title} {Spin selective charge
  recombination in chiral donor–bridge–acceptor triads},\ }\href
  {https://doi.org/10.1063/5.0150269} {\bibfield  {journal} {\bibinfo
  {journal} {The Journal of Chemical Physics}\ }\textbf {\bibinfo {volume}
  {158}},\ \bibinfo {pages} {194101} (\bibinfo {year} {2023})}\BibitemShut
  {NoStop}%
\bibitem [{\citenamefont {Chandran}\ \emph
  {et~al.}(2022{\natexlab{a}})\citenamefont {Chandran}, \citenamefont {Wu},
  \citenamefont {Teh}, \citenamefont {Waldeck},\ and\ \citenamefont
  {Subotnik}}]{chandran_electron_2022}%
  \BibitemOpen
  \bibfield  {author} {\bibinfo {author} {\bibfnamefont {S.~S.}\ \bibnamefont
  {Chandran}}, \bibinfo {author} {\bibfnamefont {Y.}~\bibnamefont {Wu}},
  \bibinfo {author} {\bibfnamefont {H.-H.}\ \bibnamefont {Teh}}, \bibinfo
  {author} {\bibfnamefont {D.~H.}\ \bibnamefont {Waldeck}},\ and\ \bibinfo
  {author} {\bibfnamefont {J.~E.}\ \bibnamefont {Subotnik}},\ }\bibfield
  {title} {\bibinfo {title} {Electron transfer and spin–orbit coupling: {Can}
  nuclear motion lead to spin selective rates?},\ }\href
  {https://doi.org/10.1063/5.0086554} {\bibfield  {journal} {\bibinfo
  {journal} {The Journal of Chemical Physics}\ }\textbf {\bibinfo {volume}
  {156}},\ \bibinfo {pages} {174113} (\bibinfo {year}
  {2022}{\natexlab{a}})}\BibitemShut {NoStop}%
\bibitem [{\citenamefont {Chandran}\ \emph
  {et~al.}(2022{\natexlab{b}})\citenamefont {Chandran}, \citenamefont {Wu},\
  and\ \citenamefont {Subotnik}}]{chandran_effect_2022}%
  \BibitemOpen
  \bibfield  {author} {\bibinfo {author} {\bibfnamefont {S.~S.}\ \bibnamefont
  {Chandran}}, \bibinfo {author} {\bibfnamefont {Y.}~\bibnamefont {Wu}},\ and\
  \bibinfo {author} {\bibfnamefont {J.~E.}\ \bibnamefont {Subotnik}},\
  }\bibfield  {title} {\bibinfo {title} {Effect of {Duschinskii} {Rotations} on
  {Spin}-{Dependent} {Electron} {Transfer} {Dynamics}},\ }\href
  {https://doi.org/10.1021/acs.jpca.2c06149} {\bibfield  {journal} {\bibinfo
  {journal} {The Journal of Physical Chemistry A}\ }\textbf {\bibinfo {volume}
  {126}},\ \bibinfo {pages} {9535} (\bibinfo {year}
  {2022}{\natexlab{b}})}\BibitemShut {NoStop}%
\bibitem [{\citenamefont {Chiesa}\ \emph {et~al.}(2024)\citenamefont {Chiesa},
  \citenamefont {Garlatti}, \citenamefont {Mezzadri}, \citenamefont {Celada},
  \citenamefont {Sessoli}, \citenamefont {Wasielewski}, \citenamefont {Bittl},
  \citenamefont {Santini},\ and\ \citenamefont
  {Carretta}}]{chiesa_many-body_2024}%
  \BibitemOpen
  \bibfield  {author} {\bibinfo {author} {\bibfnamefont {A.}~\bibnamefont
  {Chiesa}}, \bibinfo {author} {\bibfnamefont {E.}~\bibnamefont {Garlatti}},
  \bibinfo {author} {\bibfnamefont {M.}~\bibnamefont {Mezzadri}}, \bibinfo
  {author} {\bibfnamefont {L.}~\bibnamefont {Celada}}, \bibinfo {author}
  {\bibfnamefont {R.}~\bibnamefont {Sessoli}}, \bibinfo {author} {\bibfnamefont
  {M.~R.}\ \bibnamefont {Wasielewski}}, \bibinfo {author} {\bibfnamefont
  {R.}~\bibnamefont {Bittl}}, \bibinfo {author} {\bibfnamefont
  {P.}~\bibnamefont {Santini}},\ and\ \bibinfo {author} {\bibfnamefont
  {S.}~\bibnamefont {Carretta}},\ }\bibfield  {title} {\bibinfo {title}
  {Many-{Body} {Models} for {Chirality}-{Induced} {Spin} {Selectivity} in
  {Electron} {Transfer}},\ }\href
  {https://doi.org/10.1021/acs.nanolett.4c02912} {\bibfield  {journal}
  {\bibinfo  {journal} {Nano Letters}\ }\textbf {\bibinfo {volume} {24}},\
  \bibinfo {pages} {12133} (\bibinfo {year} {2024})}\BibitemShut {NoStop}%
\bibitem [{\citenamefont {Fransson}(2025)}]{fransson_chiral_2025}%
  \BibitemOpen
  \bibfield  {author} {\bibinfo {author} {\bibfnamefont {J.}~\bibnamefont
  {Fransson}},\ }\bibfield  {title} {\bibinfo {title} {Chiral {Induced} {Spin}
  {Polarized} {Electron} {Current}: {Origin} of the {Chiral} {Induced} {Spin}
  {Selectivity} {Effect}},\ }\href
  {https://doi.org/10.1021/acs.jpclett.5c00104} {\bibfield  {journal} {\bibinfo
   {journal} {The Journal of Physical Chemistry Letters}\ }\textbf {\bibinfo
  {volume} {16}},\ \bibinfo {pages} {4346} (\bibinfo {year} {2025})},\ \bibinfo
  {note} {publisher: American Chemical Society (ACS)}\BibitemShut {NoStop}%
\bibitem [{\citenamefont {Chiesa}\ \emph {et~al.}(2025)\citenamefont {Chiesa},
  \citenamefont {Privitera}, \citenamefont {Garlatti}, \citenamefont {Allodi},
  \citenamefont {Bittl}, \citenamefont {Wasielewski}, \citenamefont {Sessoli},\
  and\ \citenamefont {Carretta}}]{chiesa_chirality-induced_2025}%
  \BibitemOpen
  \bibfield  {author} {\bibinfo {author} {\bibfnamefont {A.}~\bibnamefont
  {Chiesa}}, \bibinfo {author} {\bibfnamefont {A.}~\bibnamefont {Privitera}},
  \bibinfo {author} {\bibfnamefont {E.}~\bibnamefont {Garlatti}}, \bibinfo
  {author} {\bibfnamefont {G.}~\bibnamefont {Allodi}}, \bibinfo {author}
  {\bibfnamefont {R.}~\bibnamefont {Bittl}}, \bibinfo {author} {\bibfnamefont
  {M.~R.}\ \bibnamefont {Wasielewski}}, \bibinfo {author} {\bibfnamefont
  {R.}~\bibnamefont {Sessoli}},\ and\ \bibinfo {author} {\bibfnamefont
  {S.}~\bibnamefont {Carretta}},\ }\bibfield  {title} {\bibinfo {title}
  {Chirality-{Induced} {Spin} {Selectivity} at the {Molecular} {Level}: {A}
  {Different} {Perspective} to {Understand} and {Exploit} the {Phenomenon}},\
  }\href {https://doi.org/10.1021/acs.jpclett.5c00755} {\bibfield  {journal}
  {\bibinfo  {journal} {The Journal of Physical Chemistry Letters}\ }\textbf
  {\bibinfo {volume} {16}},\ \bibinfo {pages} {5358} (\bibinfo {year}
  {2025})},\ \bibinfo {note} {publisher: American Chemical Society
  (ACS)}\BibitemShut {NoStop}%
\bibitem [{\citenamefont {Mims}\ \emph {et~al.}(2021)\citenamefont {Mims},
  \citenamefont {Herpich}, \citenamefont {Lukzen}, \citenamefont {Steiner},\
  and\ \citenamefont {Lambert}}]{mims_readout_2021}%
  \BibitemOpen
  \bibfield  {author} {\bibinfo {author} {\bibfnamefont {D.}~\bibnamefont
  {Mims}}, \bibinfo {author} {\bibfnamefont {J.}~\bibnamefont {Herpich}},
  \bibinfo {author} {\bibfnamefont {N.~N.}\ \bibnamefont {Lukzen}}, \bibinfo
  {author} {\bibfnamefont {U.~E.}\ \bibnamefont {Steiner}},\ and\ \bibinfo
  {author} {\bibfnamefont {C.}~\bibnamefont {Lambert}},\ }\bibfield  {title}
  {\bibinfo {title} {Readout of spin quantum beats in a charge-separated
  radical pair by pump-push spectroscopy},\ }\href
  {https://doi.org/10.1126/science.abl4254} {\bibfield  {journal} {\bibinfo
  {journal} {Science}\ }\textbf {\bibinfo {volume} {374}},\ \bibinfo {pages}
  {1470} (\bibinfo {year} {2021})}\BibitemShut {NoStop}%
\bibitem [{\citenamefont {Malrieu}\ \emph {et~al.}(2014)\citenamefont
  {Malrieu}, \citenamefont {Caballol}, \citenamefont {Calzado}, \citenamefont
  {De~Graaf},\ and\ \citenamefont {Guihéry}}]{malrieu_magnetic_2014}%
  \BibitemOpen
  \bibfield  {author} {\bibinfo {author} {\bibfnamefont {J.~P.}\ \bibnamefont
  {Malrieu}}, \bibinfo {author} {\bibfnamefont {R.}~\bibnamefont {Caballol}},
  \bibinfo {author} {\bibfnamefont {C.~J.}\ \bibnamefont {Calzado}}, \bibinfo
  {author} {\bibfnamefont {C.}~\bibnamefont {De~Graaf}},\ and\ \bibinfo
  {author} {\bibfnamefont {N.}~\bibnamefont {Guihéry}},\ }\bibfield  {title}
  {\bibinfo {title} {Magnetic {Interactions} in {Molecules} and {Highly}
  {Correlated} {Materials}: {Physical} {Content}, {Analytical} {Derivation},
  and {Rigorous} {Extraction} of {Magnetic} {Hamiltonians}},\ }\href
  {https://doi.org/10.1021/cr300500z} {\bibfield  {journal} {\bibinfo
  {journal} {Chemical Reviews}\ }\textbf {\bibinfo {volume} {114}},\ \bibinfo
  {pages} {429} (\bibinfo {year} {2014})}\BibitemShut {NoStop}%
\bibitem [{\citenamefont {Dalum}\ and\ \citenamefont
  {Hedegard}(2019)}]{dalum_theory_2019}%
  \BibitemOpen
  \bibfield  {author} {\bibinfo {author} {\bibfnamefont {S.}~\bibnamefont
  {Dalum}}\ and\ \bibinfo {author} {\bibfnamefont {P.}~\bibnamefont
  {Hedegard}},\ }\bibfield  {title} {\bibinfo {title} {Theory of {Chiral}
  {Induced} {Spin} {Selectivity}},\ }\href
  {https://doi.org/10.1021/acs.nanolett.9b01707} {\bibfield  {journal}
  {\bibinfo  {journal} {Nano Letters}\ }\textbf {\bibinfo {volume} {19}},\
  \bibinfo {pages} {5253} (\bibinfo {year} {2019})}\BibitemShut {NoStop}%
\bibitem [{\citenamefont {Poh}\ \emph {et~al.}(2025)\citenamefont {Poh},
  \citenamefont {Koner}, \citenamefont {Reitz},\ and\ \citenamefont
  {Yuen-Zhou}}]{poh_photoredox_2025}%
  \BibitemOpen
  \bibfield  {author} {\bibinfo {author} {\bibfnamefont {Y.~R.}\ \bibnamefont
  {Poh}}, \bibinfo {author} {\bibfnamefont {A.}~\bibnamefont {Koner}}, \bibinfo
  {author} {\bibfnamefont {M.}~\bibnamefont {Reitz}},\ and\ \bibinfo {author}
  {\bibfnamefont {J.}~\bibnamefont {Yuen-Zhou}},\ }\bibfield  {title} {\bibinfo
  {title} {Photoredox {Catalysis} with {Spin} {Magnetic} {Field} {Effects}: {A}
  {Scheme} for {Chiral} {Resolution}}\ }\href
  {https://doi.org/10.26434/chemrxiv-2025-pzqh8} {10.26434/chemrxiv-2025-pzqh8}
  (\bibinfo {year} {2025})\BibitemShut {NoStop}%
\bibitem [{\citenamefont {Fay}\ \emph {et~al.}(2019)\citenamefont {Fay},
  \citenamefont {Lindoy},\ and\ \citenamefont
  {Manolopoulos}}]{fay_electron_2019}%
  \BibitemOpen
  \bibfield  {author} {\bibinfo {author} {\bibfnamefont {T.~P.}\ \bibnamefont
  {Fay}}, \bibinfo {author} {\bibfnamefont {L.~P.}\ \bibnamefont {Lindoy}},\
  and\ \bibinfo {author} {\bibfnamefont {D.~E.}\ \bibnamefont {Manolopoulos}},\
  }\bibfield  {title} {\bibinfo {title} {Electron spin relaxation in radical
  pairs: {Beyond} the {Redfield} approximation},\ }\href
  {https://doi.org/10.1063/1.5125752} {\bibfield  {journal} {\bibinfo
  {journal} {The Journal of Chemical Physics}\ }\textbf {\bibinfo {volume}
  {151}},\ \bibinfo {pages} {154117} (\bibinfo {year} {2019})}\BibitemShut
  {NoStop}%
\bibitem [{\citenamefont {Caruso}\ \emph {et~al.}(2009)\citenamefont {Caruso},
  \citenamefont {Chin}, \citenamefont {Datta}, \citenamefont {Huelga},\ and\
  \citenamefont {Plenio}}]{caruso_highly_2009}%
  \BibitemOpen
  \bibfield  {author} {\bibinfo {author} {\bibfnamefont {F.}~\bibnamefont
  {Caruso}}, \bibinfo {author} {\bibfnamefont {A.~W.}\ \bibnamefont {Chin}},
  \bibinfo {author} {\bibfnamefont {A.}~\bibnamefont {Datta}}, \bibinfo
  {author} {\bibfnamefont {S.~F.}\ \bibnamefont {Huelga}},\ and\ \bibinfo
  {author} {\bibfnamefont {M.~B.}\ \bibnamefont {Plenio}},\ }\bibfield  {title}
  {\bibinfo {title} {Highly efficient energy excitation transfer in
  light-harvesting complexes: {The} fundamental role of noise-assisted
  transport},\ }\href {https://doi.org/10.1063/1.3223548} {\bibfield  {journal}
  {\bibinfo  {journal} {The Journal of Chemical Physics}\ }\textbf {\bibinfo
  {volume} {131}},\ \bibinfo {pages} {105106} (\bibinfo {year}
  {2009})}\BibitemShut {NoStop}%
\bibitem [{\citenamefont {Ren}\ \emph {et~al.}(2024)\citenamefont {Ren},
  \citenamefont {Shi}, \citenamefont {Feng}, \citenamefont {Xu},\ and\
  \citenamefont {Hao}}]{ren_recent_2024}%
  \BibitemOpen
  \bibfield  {author} {\bibinfo {author} {\bibfnamefont {Z.}~\bibnamefont
  {Ren}}, \bibinfo {author} {\bibfnamefont {Z.}~\bibnamefont {Shi}}, \bibinfo
  {author} {\bibfnamefont {H.}~\bibnamefont {Feng}}, \bibinfo {author}
  {\bibfnamefont {Z.}~\bibnamefont {Xu}},\ and\ \bibinfo {author}
  {\bibfnamefont {W.}~\bibnamefont {Hao}},\ }\bibfield  {title} {\bibinfo
  {title} {Recent {Progresses} of {Polarons}: {Fundamentals} and {Roles} in
  {Photocatalysis} and {Photoelectrocatalysis}},\ }\href
  {https://doi.org/10.1002/advs.202305139} {\bibfield  {journal} {\bibinfo
  {journal} {Advanced Science}\ }\textbf {\bibinfo {volume} {11}},\ \bibinfo
  {pages} {2305139} (\bibinfo {year} {2024})}\BibitemShut {NoStop}%
\bibitem [{\citenamefont {Ritson}\ and\ \citenamefont
  {Sutherland}(2013)}]{ritson_synthesis_2013}%
  \BibitemOpen
  \bibfield  {author} {\bibinfo {author} {\bibfnamefont {D.~J.}\ \bibnamefont
  {Ritson}}\ and\ \bibinfo {author} {\bibfnamefont {J.~D.}\ \bibnamefont
  {Sutherland}},\ }\bibfield  {title} {\bibinfo {title} {Synthesis of
  {Aldehydic} {Ribonucleotide} and {Amino} {Acid} {Precursors} by {Photoredox}
  {Chemistry}},\ }\href {https://doi.org/10.1002/anie.201300321} {\bibfield
  {journal} {\bibinfo  {journal} {Angewandte Chemie International Edition}\
  }\textbf {\bibinfo {volume} {52}},\ \bibinfo {pages} {5845} (\bibinfo {year}
  {2013})}\BibitemShut {NoStop}%
\bibitem [{\citenamefont {Schrupp}\ \emph {et~al.}(2005)\citenamefont
  {Schrupp}, \citenamefont {Sing}, \citenamefont {Tsunekawa}, \citenamefont
  {Fujiwara}, \citenamefont {Kasai}, \citenamefont {Sekiyama}, \citenamefont
  {Suga}, \citenamefont {Muro}, \citenamefont {Brabers},\ and\ \citenamefont
  {Claessen}}]{schrupp_high-energy_2005}%
  \BibitemOpen
  \bibfield  {author} {\bibinfo {author} {\bibfnamefont {D.}~\bibnamefont
  {Schrupp}}, \bibinfo {author} {\bibfnamefont {M.}~\bibnamefont {Sing}},
  \bibinfo {author} {\bibfnamefont {M.}~\bibnamefont {Tsunekawa}}, \bibinfo
  {author} {\bibfnamefont {H.}~\bibnamefont {Fujiwara}}, \bibinfo {author}
  {\bibfnamefont {S.}~\bibnamefont {Kasai}}, \bibinfo {author} {\bibfnamefont
  {A.}~\bibnamefont {Sekiyama}}, \bibinfo {author} {\bibfnamefont
  {S.}~\bibnamefont {Suga}}, \bibinfo {author} {\bibfnamefont {T.}~\bibnamefont
  {Muro}}, \bibinfo {author} {\bibfnamefont {V.~A.~M.}\ \bibnamefont
  {Brabers}},\ and\ \bibinfo {author} {\bibfnamefont {R.}~\bibnamefont
  {Claessen}},\ }\bibfield  {title} {\bibinfo {title} {High-energy
  photoemission on {Fe}$_{\textrm{3}}$ {O}$_{\textrm{4}}$ : {Small} polaron
  physics and the {Verwey} transition},\ }\href
  {https://doi.org/10.1209/epl/i2005-10045-y} {\bibfield  {journal} {\bibinfo
  {journal} {Europhysics Letters (EPL)}\ }\textbf {\bibinfo {volume} {70}},\
  \bibinfo {pages} {789} (\bibinfo {year} {2005})}\BibitemShut {NoStop}%
\bibitem [{\citenamefont {Morse}\ \emph {et~al.}(2018)\citenamefont {Morse},
  \citenamefont {Nguyen}, \citenamefont {Cruz},\ and\ \citenamefont
  {Nicewicz}}]{morse_enantioselective_2018}%
  \BibitemOpen
  \bibfield  {author} {\bibinfo {author} {\bibfnamefont {P.~D.}\ \bibnamefont
  {Morse}}, \bibinfo {author} {\bibfnamefont {T.~M.}\ \bibnamefont {Nguyen}},
  \bibinfo {author} {\bibfnamefont {C.~L.}\ \bibnamefont {Cruz}},\ and\
  \bibinfo {author} {\bibfnamefont {D.~A.}\ \bibnamefont {Nicewicz}},\
  }\bibfield  {title} {\bibinfo {title} {Enantioselective counter-anions in
  photoredox catalysis: {The} asymmetric cation radical {Diels}-{Alder}
  reaction},\ }\href {https://doi.org/10.1016/j.tet.2018.03.052} {\bibfield
  {journal} {\bibinfo  {journal} {Tetrahedron}\ }\textbf {\bibinfo {volume}
  {74}},\ \bibinfo {pages} {3266} (\bibinfo {year} {2018})}\BibitemShut
  {NoStop}%
\bibitem [{\citenamefont {Brugh}\ and\ \citenamefont
  {Forbes}(2020)}]{brugh_anomalous_2020}%
  \BibitemOpen
  \bibfield  {author} {\bibinfo {author} {\bibfnamefont {A.~M.}\ \bibnamefont
  {Brugh}}\ and\ \bibinfo {author} {\bibfnamefont {M.~D.~E.}\ \bibnamefont
  {Forbes}},\ }\bibfield  {title} {\bibinfo {title} {Anomalous chemically
  induced electron spin polarization in proton-coupled electron transfer
  reactions: insight into radical pair dynamics},\ }\href
  {https://doi.org/10.1039/D0SC02691C} {\bibfield  {journal} {\bibinfo
  {journal} {Chemical Science}\ }\textbf {\bibinfo {volume} {11}},\ \bibinfo
  {pages} {6268} (\bibinfo {year} {2020})}\BibitemShut {NoStop}%
\bibitem [{\citenamefont {Zhang}\ \emph {et~al.}(2014)\citenamefont {Zhang},
  \citenamefont {Samanta}, \citenamefont {Rosen},\ and\ \citenamefont
  {Percec}}]{zhang_single_2014}%
  \BibitemOpen
  \bibfield  {author} {\bibinfo {author} {\bibfnamefont {N.}~\bibnamefont
  {Zhang}}, \bibinfo {author} {\bibfnamefont {S.~R.}\ \bibnamefont {Samanta}},
  \bibinfo {author} {\bibfnamefont {B.~M.}\ \bibnamefont {Rosen}},\ and\
  \bibinfo {author} {\bibfnamefont {V.}~\bibnamefont {Percec}},\ }\bibfield
  {title} {\bibinfo {title} {Single {Electron} {Transfer} in {Radical} {Ion}
  and {Radical}-{Mediated} {Organic}, {Materials} and {Polymer} {Synthesis}},\
  }\href {https://doi.org/10.1021/cr400689s} {\bibfield  {journal} {\bibinfo
  {journal} {Chemical Reviews}\ }\textbf {\bibinfo {volume} {114}},\ \bibinfo
  {pages} {5848} (\bibinfo {year} {2014})}\BibitemShut {NoStop}%
\bibitem [{\citenamefont {Kattnig}\ and\ \citenamefont
  {Hore}(2017)}]{kattnig_sensitivity_2017}%
  \BibitemOpen
  \bibfield  {author} {\bibinfo {author} {\bibfnamefont {D.~R.}\ \bibnamefont
  {Kattnig}}\ and\ \bibinfo {author} {\bibfnamefont {P.~J.}\ \bibnamefont
  {Hore}},\ }\bibfield  {title} {\bibinfo {title} {The sensitivity of a radical
  pair compass magnetoreceptor can be significantly amplified by radical
  scavengers},\ }\href {https://doi.org/10.1038/s41598-017-09914-7} {\bibfield
  {journal} {\bibinfo  {journal} {Scientific Reports}\ }\textbf {\bibinfo
  {volume} {7}},\ \bibinfo {pages} {11640} (\bibinfo {year}
  {2017})}\BibitemShut {NoStop}%
\bibitem [{\citenamefont {Henbest}\ \emph {et~al.}(2004)\citenamefont
  {Henbest}, \citenamefont {Kukura}, \citenamefont {Rodgers}, \citenamefont
  {Hore},\ and\ \citenamefont {Timmel}}]{henbest_radio_2004}%
  \BibitemOpen
  \bibfield  {author} {\bibinfo {author} {\bibfnamefont {K.~B.}\ \bibnamefont
  {Henbest}}, \bibinfo {author} {\bibfnamefont {P.}~\bibnamefont {Kukura}},
  \bibinfo {author} {\bibfnamefont {C.~T.}\ \bibnamefont {Rodgers}}, \bibinfo
  {author} {\bibfnamefont {P.~J.}\ \bibnamefont {Hore}},\ and\ \bibinfo
  {author} {\bibfnamefont {C.~R.}\ \bibnamefont {Timmel}},\ }\bibfield  {title}
  {\bibinfo {title} {Radio {Frequency} {Magnetic} {Field} {Effects} on a
  {Radical} {Recombination} {Reaction}: {A} {Diagnostic} {Test} for the
  {Radical} {Pair} {Mechanism}},\ }\href {https://doi.org/10.1021/ja048220q}
  {\bibfield  {journal} {\bibinfo  {journal} {Journal of the American Chemical
  Society}\ }\textbf {\bibinfo {volume} {126}},\ \bibinfo {pages} {8102}
  (\bibinfo {year} {2004})}\BibitemShut {NoStop}%
\bibitem [{\citenamefont {Buchachenko}(2001)}]{buchachenko_magnetic_2001}%
  \BibitemOpen
  \bibfield  {author} {\bibinfo {author} {\bibfnamefont {A.~L.}\ \bibnamefont
  {Buchachenko}},\ }\bibfield  {title} {\bibinfo {title} {Magnetic {Isotope}
  {Effect}: {Nuclear} {Spin} {Control} of {Chemical} {Reactions}},\ }\href
  {https://doi.org/10.1021/jp011261d} {\bibfield  {journal} {\bibinfo
  {journal} {The Journal of Physical Chemistry A}\ }\textbf {\bibinfo {volume}
  {105}},\ \bibinfo {pages} {9995} (\bibinfo {year} {2001})}\BibitemShut
  {NoStop}%
\bibitem [{\citenamefont {Lawrence}\ and\ \citenamefont
  {Manolopoulos}(2020)}]{lawrence_path_2020}%
  \BibitemOpen
  \bibfield  {author} {\bibinfo {author} {\bibfnamefont {J.~E.}\ \bibnamefont
  {Lawrence}}\ and\ \bibinfo {author} {\bibfnamefont {D.~E.}\ \bibnamefont
  {Manolopoulos}},\ }\bibfield  {title} {\bibinfo {title} {Path integral
  methods for reaction rates in complex systems},\ }\href
  {https://doi.org/10.1039/C9FD00084D} {\bibfield  {journal} {\bibinfo
  {journal} {Faraday Discussions}\ }\textbf {\bibinfo {volume} {221}},\
  \bibinfo {pages} {9} (\bibinfo {year} {2020})}\BibitemShut {NoStop}%
\bibitem [{\citenamefont {Blumberger}(2015)}]{blumberger_recent_2015}%
  \BibitemOpen
  \bibfield  {author} {\bibinfo {author} {\bibfnamefont {J.}~\bibnamefont
  {Blumberger}},\ }\bibfield  {title} {\bibinfo {title} {Recent {Advances} in
  the {Theory} and {Molecular} {Simulation} of {Biological} {Electron}
  {Transfer} {Reactions}},\ }\href
  {https://doi.org/10.1021/acs.chemrev.5b00298} {\bibfield  {journal} {\bibinfo
   {journal} {Chemical Reviews}\ }\textbf {\bibinfo {volume} {115}},\ \bibinfo
  {pages} {11191} (\bibinfo {year} {2015})}\BibitemShut {NoStop}%
\bibitem [{\citenamefont {Fay}\ \emph {et~al.}(2018)\citenamefont {Fay},
  \citenamefont {Lindoy},\ and\ \citenamefont
  {Manolopoulos}}]{fay_spin-selective_2018}%
  \BibitemOpen
  \bibfield  {author} {\bibinfo {author} {\bibfnamefont {T.~P.}\ \bibnamefont
  {Fay}}, \bibinfo {author} {\bibfnamefont {L.~P.}\ \bibnamefont {Lindoy}},\
  and\ \bibinfo {author} {\bibfnamefont {D.~E.}\ \bibnamefont {Manolopoulos}},\
  }\bibfield  {title} {\bibinfo {title} {Spin-selective electron transfer
  reactions of radical pairs: {Beyond} the {Haberkorn} master equation},\
  }\href {https://doi.org/10.1063/1.5041520} {\bibfield  {journal} {\bibinfo
  {journal} {The Journal of Chemical Physics}\ }\textbf {\bibinfo {volume}
  {149}},\ \bibinfo {pages} {064107} (\bibinfo {year} {2018})}\BibitemShut
  {NoStop}%
\end{thebibliography}%


\begin{thebibliography}{16}%
\makeatletter
\providecommand \@ifxundefined [1]{%
 \@ifx{#1\undefined}
}%
\providecommand \@ifnum [1]{%
 \ifnum #1\expandafter \@firstoftwo
 \else \expandafter \@secondoftwo
 \fi
}%
\providecommand \@ifx [1]{%
 \ifx #1\expandafter \@firstoftwo
 \else \expandafter \@secondoftwo
 \fi
}%
\providecommand \natexlab [1]{#1}%
\providecommand \enquote  [1]{``#1''}%
\providecommand \bibnamefont  [1]{#1}%
\providecommand \bibfnamefont [1]{#1}%
\providecommand \citenamefont [1]{#1}%
\providecommand \href@noop [0]{\@secondoftwo}%
\providecommand \href [0]{\begingroup \@sanitize@url \@href}%
\providecommand \@href[1]{\@@startlink{#1}\@@href}%
\providecommand \@@href[1]{\endgroup#1\@@endlink}%
\providecommand \@sanitize@url [0]{\catcode `\\12\catcode `\$12\catcode
  `\&12\catcode `\#12\catcode `\^12\catcode `\_12\catcode `\%12\relax}%
\providecommand \@@startlink[1]{}%
\providecommand \@@endlink[0]{}%
\providecommand \url  [0]{\begingroup\@sanitize@url \@url }%
\providecommand \@url [1]{\endgroup\@href {#1}{\urlprefix }}%
\providecommand \urlprefix  [0]{URL }%
\providecommand \Eprint [0]{\href }%
\providecommand \doibase [0]{https://doi.org/}%
\providecommand \selectlanguage [0]{\@gobble}%
\providecommand \bibinfo  [0]{\@secondoftwo}%
\providecommand \bibfield  [0]{\@secondoftwo}%
\providecommand \translation [1]{[#1]}%
\providecommand \BibitemOpen [0]{}%
\providecommand \bibitemStop [0]{}%
\providecommand \bibitemNoStop [0]{.\EOS\space}%
\providecommand \EOS [0]{\spacefactor3000\relax}%
\providecommand \BibitemShut  [1]{\csname bibitem#1\endcsname}%
\let\auto@bib@innerbib\@empty
\bibitem [{\citenamefont {Fay}, \citenamefont {Lindoy},\ and\ \citenamefont
  {Manolopoulos}(2018)}]{fay_spin-selective_2018}%
  \BibitemOpen
  \bibfield  {author} {\bibinfo {author} {\bibfnamefont {T.~P.}\ \bibnamefont
  {Fay}}, \bibinfo {author} {\bibfnamefont {L.~P.}\ \bibnamefont {Lindoy}},\
  and\ \bibinfo {author} {\bibfnamefont {D.~E.}\ \bibnamefont {Manolopoulos}},\
  }\bibfield  {title} {\enquote {\bibinfo {title} {Spin-selective electron
  transfer reactions of radical pairs: {Beyond} the {Haberkorn} master
  equation},}\ }\href {https://doi.org/10.1063/1.5041520} {\bibfield  {journal}
  {\bibinfo  {journal} {The Journal of Chemical Physics}\ }\textbf {\bibinfo
  {volume} {149}},\ \bibinfo {pages} {064107} (\bibinfo {year}
  {2018})}\BibitemShut {NoStop}%
\bibitem [{\citenamefont {Fay}(2021)}]{fay_chirality-induced_2021}%
  \BibitemOpen
  \bibfield  {author} {\bibinfo {author} {\bibfnamefont {T.~P.}\ \bibnamefont
  {Fay}},\ }\bibfield  {title} {\enquote {\bibinfo {title} {Chirality-{Induced}
  {Spin} {Coherence} in {Electron} {Transfer} {Reactions}},}\ }\href
  {https://doi.org/10.1021/acs.jpclett.1c00009} {\bibfield  {journal} {\bibinfo
   {journal} {The Journal of Physical Chemistry Letters}\ }\textbf {\bibinfo
  {volume} {12}},\ \bibinfo {pages} {1407--1412} (\bibinfo {year}
  {2021})}\BibitemShut {NoStop}%
\bibitem [{\citenamefont {Steiner}\ and\ \citenamefont
  {Ulrich}(1989)}]{steiner_magnetic_1989}%
  \BibitemOpen
  \bibfield  {author} {\bibinfo {author} {\bibfnamefont {U.~E.}\ \bibnamefont
  {Steiner}}\ and\ \bibinfo {author} {\bibfnamefont {T.}~\bibnamefont
  {Ulrich}},\ }\bibfield  {title} {\enquote {\bibinfo {title} {Magnetic field
  effects in chemical kinetics and related phenomena},}\ }\href
  {https://doi.org/10.1021/cr00091a003} {\bibfield  {journal} {\bibinfo
  {journal} {Chemical Reviews}\ }\textbf {\bibinfo {volume} {89}},\ \bibinfo
  {pages} {51--147} (\bibinfo {year} {1989})}\BibitemShut {NoStop}%
\bibitem [{\citenamefont {Fay}, \citenamefont {Lindoy},\ and\ \citenamefont
  {Manolopoulos}(2019)}]{fay_electron_2019}%
  \BibitemOpen
  \bibfield  {author} {\bibinfo {author} {\bibfnamefont {T.~P.}\ \bibnamefont
  {Fay}}, \bibinfo {author} {\bibfnamefont {L.~P.}\ \bibnamefont {Lindoy}},\
  and\ \bibinfo {author} {\bibfnamefont {D.~E.}\ \bibnamefont {Manolopoulos}},\
  }\bibfield  {title} {\enquote {\bibinfo {title} {Electron spin relaxation in
  radical pairs: {Beyond} the {Redfield} approximation},}\ }\href
  {https://doi.org/10.1063/1.5125752} {\bibfield  {journal} {\bibinfo
  {journal} {The Journal of Chemical Physics}\ }\textbf {\bibinfo {volume}
  {151}},\ \bibinfo {pages} {154117} (\bibinfo {year} {2019})}\BibitemShut
  {NoStop}%
\bibitem [{\citenamefont {Lawrence}\ and\ \citenamefont
  {Manolopoulos}(2020)}]{lawrence_path_2020}%
  \BibitemOpen
  \bibfield  {author} {\bibinfo {author} {\bibfnamefont {J.~E.}\ \bibnamefont
  {Lawrence}}\ and\ \bibinfo {author} {\bibfnamefont {D.~E.}\ \bibnamefont
  {Manolopoulos}},\ }\bibfield  {title} {\enquote {\bibinfo {title} {Path
  integral methods for reaction rates in complex systems},}\ }\href
  {https://doi.org/10.1039/C9FD00084D} {\bibfield  {journal} {\bibinfo
  {journal} {Faraday Discussions}\ }\textbf {\bibinfo {volume} {221}},\
  \bibinfo {pages} {9--29} (\bibinfo {year} {2020})}\BibitemShut {NoStop}%
\bibitem [{\citenamefont {Blumberger}(2015)}]{blumberger_recent_2015}%
  \BibitemOpen
  \bibfield  {author} {\bibinfo {author} {\bibfnamefont {J.}~\bibnamefont
  {Blumberger}},\ }\bibfield  {title} {\enquote {\bibinfo {title} {Recent
  {Advances} in the {Theory} and {Molecular} {Simulation} of {Biological}
  {Electron} {Transfer} {Reactions}},}\ }\href
  {https://doi.org/10.1021/acs.chemrev.5b00298} {\bibfield  {journal} {\bibinfo
   {journal} {Chemical Reviews}\ }\textbf {\bibinfo {volume} {115}},\ \bibinfo
  {pages} {11191--11238} (\bibinfo {year} {2015})}\BibitemShut {NoStop}%
\bibitem [{\citenamefont {Malrieu}\ \emph {et~al.}(2014)\citenamefont
  {Malrieu}, \citenamefont {Caballol}, \citenamefont {Calzado}, \citenamefont
  {De~Graaf},\ and\ \citenamefont {Guihéry}}]{malrieu_magnetic_2014}%
  \BibitemOpen
  \bibfield  {author} {\bibinfo {author} {\bibfnamefont {J.~P.}\ \bibnamefont
  {Malrieu}}, \bibinfo {author} {\bibfnamefont {R.}~\bibnamefont {Caballol}},
  \bibinfo {author} {\bibfnamefont {C.~J.}\ \bibnamefont {Calzado}}, \bibinfo
  {author} {\bibfnamefont {C.}~\bibnamefont {De~Graaf}},\ and\ \bibinfo
  {author} {\bibfnamefont {N.}~\bibnamefont {Guihéry}},\ }\bibfield  {title}
  {\enquote {\bibinfo {title} {Magnetic {Interactions} in {Molecules} and
  {Highly} {Correlated} {Materials}: {Physical} {Content}, {Analytical}
  {Derivation}, and {Rigorous} {Extraction} of {Magnetic} {Hamiltonians}},}\
  }\href {https://doi.org/10.1021/cr300500z} {\bibfield  {journal} {\bibinfo
  {journal} {Chemical Reviews}\ }\textbf {\bibinfo {volume} {114}},\ \bibinfo
  {pages} {429--492} (\bibinfo {year} {2014})}\BibitemShut {NoStop}%
\bibitem [{\citenamefont {Luo}(2024)}]{luo_sensitivity_2024}%
  \BibitemOpen
  \bibfield  {author} {\bibinfo {author} {\bibfnamefont {J.}~\bibnamefont
  {Luo}},\ }\bibfield  {title} {\enquote {\bibinfo {title} {Sensitivity
  enhancement of radical-pair magnetoreceptors as a result of spin
  decoherence},}\ }\href {https://doi.org/10.1063/5.0182172} {\bibfield
  {journal} {\bibinfo  {journal} {The Journal of Chemical Physics}\ }\textbf
  {\bibinfo {volume} {160}},\ \bibinfo {pages} {074306} (\bibinfo {year}
  {2024})}\BibitemShut {NoStop}%
\bibitem [{\citenamefont {Inc.}()}]{Mathematica}%
  \BibitemOpen
  \bibfield  {author} {\bibinfo {author} {\bibfnamefont {W.~R.}\ \bibnamefont
  {Inc.}},\ }\href {https://www.wolfram.com/mathematica} {\enquote {\bibinfo
  {title} {Mathematica, {V}ersion 14.1},}\ }\bibinfo {note} {Champaign, IL,
  2024}\BibitemShut {NoStop}%
\bibitem [{\citenamefont {Metzger}\ \emph {et~al.}(2021)\citenamefont
  {Metzger}, \citenamefont {Siam}, \citenamefont {Kolodny}, \citenamefont
  {Goren}, \citenamefont {Sukenik}, \citenamefont {Yochelis}, \citenamefont
  {Abu-Reziq}, \citenamefont {Avnir},\ and\ \citenamefont
  {Paltiel}}]{metzger_dynamic_2021}%
  \BibitemOpen
  \bibfield  {author} {\bibinfo {author} {\bibfnamefont {T.~S.}\ \bibnamefont
  {Metzger}}, \bibinfo {author} {\bibfnamefont {R.}~\bibnamefont {Siam}},
  \bibinfo {author} {\bibfnamefont {Y.}~\bibnamefont {Kolodny}}, \bibinfo
  {author} {\bibfnamefont {N.}~\bibnamefont {Goren}}, \bibinfo {author}
  {\bibfnamefont {N.}~\bibnamefont {Sukenik}}, \bibinfo {author} {\bibfnamefont
  {S.}~\bibnamefont {Yochelis}}, \bibinfo {author} {\bibfnamefont
  {R.}~\bibnamefont {Abu-Reziq}}, \bibinfo {author} {\bibfnamefont
  {D.}~\bibnamefont {Avnir}},\ and\ \bibinfo {author} {\bibfnamefont
  {Y.}~\bibnamefont {Paltiel}},\ }\bibfield  {title} {\enquote {\bibinfo
  {title} {Dynamic {Spin}-{Controlled} {Enantioselective} {Catalytic} {Chiral}
  {Reactions}},}\ }\href {https://doi.org/10.1021/acs.jpclett.1c01518}
  {\bibfield  {journal} {\bibinfo  {journal} {The Journal of Physical Chemistry
  Letters}\ }\textbf {\bibinfo {volume} {12}},\ \bibinfo {pages} {5469--5472}
  (\bibinfo {year} {2021})}\BibitemShut {NoStop}%
\bibitem [{\citenamefont {Morse}\ \emph {et~al.}(2018)\citenamefont {Morse},
  \citenamefont {Nguyen}, \citenamefont {Cruz},\ and\ \citenamefont
  {Nicewicz}}]{morse_enantioselective_2018}%
  \BibitemOpen
  \bibfield  {author} {\bibinfo {author} {\bibfnamefont {P.~D.}\ \bibnamefont
  {Morse}}, \bibinfo {author} {\bibfnamefont {T.~M.}\ \bibnamefont {Nguyen}},
  \bibinfo {author} {\bibfnamefont {C.~L.}\ \bibnamefont {Cruz}},\ and\
  \bibinfo {author} {\bibfnamefont {D.~A.}\ \bibnamefont {Nicewicz}},\
  }\bibfield  {title} {\enquote {\bibinfo {title} {Enantioselective
  counter-anions in photoredox catalysis: {The} asymmetric cation radical
  {Diels}-{Alder} reaction},}\ }\href
  {https://doi.org/10.1016/j.tet.2018.03.052} {\bibfield  {journal} {\bibinfo
  {journal} {Tetrahedron}\ }\textbf {\bibinfo {volume} {74}},\ \bibinfo {pages}
  {3266--3272} (\bibinfo {year} {2018})}\BibitemShut {NoStop}%
\bibitem [{\citenamefont {Brugh}\ and\ \citenamefont
  {Forbes}(2020)}]{brugh_anomalous_2020}%
  \BibitemOpen
  \bibfield  {author} {\bibinfo {author} {\bibfnamefont {A.~M.}\ \bibnamefont
  {Brugh}}\ and\ \bibinfo {author} {\bibfnamefont {M.~D.~E.}\ \bibnamefont
  {Forbes}},\ }\bibfield  {title} {\enquote {\bibinfo {title} {Anomalous
  chemically induced electron spin polarization in proton-coupled electron
  transfer reactions: insight into radical pair dynamics},}\ }\href
  {https://doi.org/10.1039/D0SC02691C} {\bibfield  {journal} {\bibinfo
  {journal} {Chemical Science}\ }\textbf {\bibinfo {volume} {11}},\ \bibinfo
  {pages} {6268--6274} (\bibinfo {year} {2020})}\BibitemShut {NoStop}%
\bibitem [{\citenamefont {Zhang}\ \emph {et~al.}(2014)\citenamefont {Zhang},
  \citenamefont {Samanta}, \citenamefont {Rosen},\ and\ \citenamefont
  {Percec}}]{zhang_single_2014}%
  \BibitemOpen
  \bibfield  {author} {\bibinfo {author} {\bibfnamefont {N.}~\bibnamefont
  {Zhang}}, \bibinfo {author} {\bibfnamefont {S.~R.}\ \bibnamefont {Samanta}},
  \bibinfo {author} {\bibfnamefont {B.~M.}\ \bibnamefont {Rosen}},\ and\
  \bibinfo {author} {\bibfnamefont {V.}~\bibnamefont {Percec}},\ }\bibfield
  {title} {\enquote {\bibinfo {title} {Single {Electron} {Transfer} in
  {Radical} {Ion} and {Radical}-{Mediated} {Organic}, {Materials} and {Polymer}
  {Synthesis}},}\ }\href {https://doi.org/10.1021/cr400689s} {\bibfield
  {journal} {\bibinfo  {journal} {Chemical Reviews}\ }\textbf {\bibinfo
  {volume} {114}},\ \bibinfo {pages} {5848--5958} (\bibinfo {year}
  {2014})}\BibitemShut {NoStop}%
\bibitem [{\citenamefont {Kattnig}\ and\ \citenamefont
  {Hore}(2017)}]{kattnig_sensitivity_2017}%
  \BibitemOpen
  \bibfield  {author} {\bibinfo {author} {\bibfnamefont {D.~R.}\ \bibnamefont
  {Kattnig}}\ and\ \bibinfo {author} {\bibfnamefont {P.~J.}\ \bibnamefont
  {Hore}},\ }\bibfield  {title} {\enquote {\bibinfo {title} {The sensitivity of
  a radical pair compass magnetoreceptor can be significantly amplified by
  radical scavengers},}\ }\href {https://doi.org/10.1038/s41598-017-09914-7}
  {\bibfield  {journal} {\bibinfo  {journal} {Scientific Reports}\ }\textbf
  {\bibinfo {volume} {7}},\ \bibinfo {pages} {11640} (\bibinfo {year}
  {2017})}\BibitemShut {NoStop}%
\bibitem [{\citenamefont {Henbest}\ \emph {et~al.}(2004)\citenamefont
  {Henbest}, \citenamefont {Kukura}, \citenamefont {Rodgers}, \citenamefont
  {Hore},\ and\ \citenamefont {Timmel}}]{henbest_radio_2004}%
  \BibitemOpen
  \bibfield  {author} {\bibinfo {author} {\bibfnamefont {K.~B.}\ \bibnamefont
  {Henbest}}, \bibinfo {author} {\bibfnamefont {P.}~\bibnamefont {Kukura}},
  \bibinfo {author} {\bibfnamefont {C.~T.}\ \bibnamefont {Rodgers}}, \bibinfo
  {author} {\bibfnamefont {P.~J.}\ \bibnamefont {Hore}},\ and\ \bibinfo
  {author} {\bibfnamefont {C.~R.}\ \bibnamefont {Timmel}},\ }\bibfield  {title}
  {\enquote {\bibinfo {title} {Radio {Frequency} {Magnetic} {Field} {Effects}
  on a {Radical} {Recombination} {Reaction}: {A} {Diagnostic} {Test} for the
  {Radical} {Pair} {Mechanism}},}\ }\href {https://doi.org/10.1021/ja048220q}
  {\bibfield  {journal} {\bibinfo  {journal} {Journal of the American Chemical
  Society}\ }\textbf {\bibinfo {volume} {126}},\ \bibinfo {pages} {8102--8103}
  (\bibinfo {year} {2004})}\BibitemShut {NoStop}%
\bibitem [{\citenamefont {Buchachenko}(2001)}]{buchachenko_magnetic_2001}%
  \BibitemOpen
  \bibfield  {author} {\bibinfo {author} {\bibfnamefont {A.~L.}\ \bibnamefont
  {Buchachenko}},\ }\bibfield  {title} {\enquote {\bibinfo {title} {Magnetic
  {Isotope} {Effect}: {Nuclear} {Spin} {Control} of {Chemical} {Reactions}},}\
  }\href {https://doi.org/10.1021/jp011261d} {\bibfield  {journal} {\bibinfo
  {journal} {The Journal of Physical Chemistry A}\ }\textbf {\bibinfo {volume}
  {105}},\ \bibinfo {pages} {9995--10011} (\bibinfo {year} {2001})}\BibitemShut
  {NoStop}%
\end{thebibliography}%

\end{document}


\setlength{\abovedisplayskip}{4pt}
	\setlength{\belowdisplayskip}{4pt}
	
		{
		\makeatletter
		\def\frontmatter@thefootnote{%
			\altaffilletter@sw{\@fnsymbol}{\@fnsymbol}{\csname c@\@mpfn\endcsname}%
		}%
		\makeatother
	\title{Supporting Information to Enantioselective radical reactions can be induced by electron spin polarization: A quantum mechanism for Nature's emergent homochirality?}
	\author{Thomas P. Fay}
	\email[Corresponding author:\ ]{thomaspfay@ucla.edu}
	\affiliation{Department of Chemistry and Biochemistry, University of California, Los Angeles, California 90095, United States}

	\maketitle
	\tableofcontents
	
	\section{Quantitative modelling of asymmetric radical pair redox reactions}\label{sec-quant}
	\vspace{-10pt}
	The explanation in the main text outlines qualitatively how spin-polarisation and spin-orbit coupling mediated electron transfer reactions of the \ce{[A^$\bullet $B^$\bullet$]} radical pair gives rise to enantiomeric excess in the final products. This simplified explanation however ignores the fact that spin-selective reactions also change coherent spin dynamics of the radical pair state, and it also ignores the effective spin-orbit interaction between radicals that will also perturb the coherent dynamics.\cite{fay_spin-selective_2018,fay_chirality-induced_2021} In order to develop a more complete and quantitative theory of spin-controlled enantioselectivity in radical pair reactions, we describe the system in terms of spin density operators of the \ce{[A^$\bullet $B^$\bullet$]_{$R/S$}} intermediates, denoted $\hat{\sigma}_{R}(t)$ and $\hat{\sigma}_S(t)$ respectively.\cite{fay_chirality-induced_2021,fay_spin-selective_2018} These spin-density operators encode the ensemble averaged spin-populations and coherences in each state, and the trace of each spin density operator corresponds to the population of the \ce{[A^$\bullet $B^$\bullet$]_{$R/S$}} state $p_{R/S}(t) = \Tr[\hat{\sigma}_{R/S}(t)]$. Following from the well-established theory of these spin-density operators,\cite{steiner_magnetic_1989,fay_spin-selective_2018} they evolve under the following coupled Stochastic Liouville equations,\cite{fay_spin-selective_2018,fay_electron_2019}
		\begin{align}\label{eq-sigma-t}
			\begin{split}
				\dv{t}\hat{\sigma}_R(t) \!&=\! -\frac{i}{\hbar} [\hat{H}_+,\hat{\sigma}_R(t)] \!-\! \{\hat{K}_+,\hat{\sigma}_R(t)\} \!-\! k_{\mathrm{f}} \hat{\sigma}_R(t) \!+\! k_{\mathrm{f}}\hat{\sigma}_S(t)\\
				\dv{t}\hat{\sigma}_S(t) \!&=\! -\frac{i}{\hbar} [\hat{H}_-,\hat{\sigma}_S(t)] \!-\! \{\hat{K}_-,\hat{\sigma}_S(t)\} \!-\! k_{\mathrm{f}} \hat{\sigma}_S(t) \!+\! k_{\mathrm{f}}\hat{\sigma}_R(t).
			\end{split}
		\end{align}
		The $-(i/\hbar)[\hat{H}_\pm,\cdot]$ term describes the coherent spin evolution in each of the \ce{[A^$\bullet $B^$\bullet$]_{$R/S$}} states, where the spin Hamiltonians are
		\begin{align}
			\hat{H}_{\pm} = 2\hbar J \dyad{\mathrm{S}} + \hbar\epsilon\dyad{\psi_{\pm\theta}}
		\end{align}
		where $J$ is the exchange (scalar) coupling between electron spins and $\epsilon$ is the effective spin-orbit coupling given approximately by $\hbar\epsilon \approx (V_{\mathrm{AB}}^2 + (\Lambda_{\mathrm{AB}}/2)^2)/\Delta E $ where $\Delta E$ is the vertical energy gap between the product and radical pair states at the equilibrium radical pair geometry. The $-\{\hat{K}_\pm,\cdot\}$ describes the spin selective reactions of the \ce{[A^$\bullet $B^$\bullet$]_{$R/S$}} states, where $\{ \cdot , \cdot \}$ denotes the anti-commutator and the reaction operators are\cite{fay_chirality-induced_2021}
		\begin{align}
			\hat{K}_{\pm} = \frac{k_\mathrm{r}}{2}\dyad{\psi_{\pm\theta}}.
		\end{align}
		$k_\mathrm{r}$ is the electron transfer rate from the $\psi_{\pm\theta}$ state of the \ce{[A^$\bullet $B^$\bullet$]_{$R/S$}} radical pair to the singlet product \ce{{}^SP_{$R/S$}}. This rate must be identical for both enantiomers. The last two terms account for the interconversion between the $R$ and $S$ precursor radical pair states, where $k_\mathrm{f}$ is the interconversion rate constant. This scheme for the radical pair dynamics is illustrated in Fig.~\ref{fig-scheme}. It should be noted that whilst this equation may appear phenomenological, it can be rigorously derived for a general fully atomistic description of the reaction through the framework of open quantum systems,\cite{fay_spin-selective_2018,fay_chirality-induced_2021} and all parameters appearing can be in principle be calculated using \textit{ab initio} quantum dynamics methods,\cite{lawrence_path_2020,blumberger_recent_2015,malrieu_magnetic_2014} although in practice additional approximations would inevitably be needed to calculate parameters for a real system. Furthermore this scheme ignores the effects of spin-relaxation (apart from those induced by $R/S$ interconversion changing $\hat{H}_\pm$ and $\hat{K}_{\pm}$), which is a reasonable assumption if the spin-relaxation time is slow compared to the other time-scales in the system. Numerical treatment of spin-relaxation can be achieved through various techniques.\cite{steiner_magnetic_1989,fay_electron_2019} It should be noted that spin-relaxation can sometimes enhance spin-selectivity in radical pair systems, so inclusion of these effects may enhance the enantioselectivity described here.\cite{luo_sensitivity_2024} 
		
		From this model we can compute the time-dependent yield of the $R$ and $S$ products as
		\begin{subequations}\label{eq-yields}
			\begin{align}
				\phi_{R}(t) &= \int_0^t 2\Tr[\hat{K}_+\hat{\sigma}_R(\tau)]\dd{\tau} \\
				\phi_{S}(t) &= \int_0^t 2\Tr[\hat{K}_-\hat{\sigma}_S(\tau)]\dd{\tau}.
			\end{align}
		\end{subequations}
		The final yields of the $R$ and $S$ products are defined as $\Phi_{R/S} = 
		\phi_{R/S}(t=\infty)$. The initial conditions for the reduced density operators are simply $\hat{\sigma}_{R}(0) = \hat{\sigma}_{S}(0) = \hat{\sigma}_0/2$, since the $R$ and $S$ precursor states are equally populated before any reaction occurs. If the degree of polarisation of \ce{A^{$\bullet$}} radical is $p_{\mathrm{A}}$, then the initial spin density operator $\hat{\sigma}_0$ can be written as
		\begin{align}
			\begin{split}
				\hat{\sigma}_0 &= \frac{1+p_\mathrm{A}}{4}(\dyad{\mathrm{T}_{+1}} + \dyad{\uparrow_{\mathrm{A}} \downarrow_{\mathrm{B}} }) \\
				&+ \frac{1-p_\mathrm{A}}{4}(\dyad{\mathrm{T}_{-1}} + \dyad{\downarrow_{\mathrm{A}} \uparrow_{\mathrm{B}} }),
			\end{split}
		\end{align}
		where $p_\mathrm{A}=+1$ corresponds to a fully up spin-polarised initial state and $p_\mathrm{A}=-1$ corresponds to a fully down spin-polarised initial state. The final yield of the $R$ and $S$ products can be obtained by integrating Eq.~\eqref{eq-sigma-t} from $t=0$ to $\infty$ (noting that $\hat{\sigma}_{R/S}(\infty) = 0$ except for the $\mathrm{T}_{+1}$ fraction, which does not react) to yield a set of linear equations for the integrated density operators. These integrated density operators can then be substituted into Eq.~\eqref{eq-yields} to obtain the final yields.\cite{steiner_magnetic_1989} This is straightforward to do with standard symbolic computation packages such as Mathematica.\cite{Mathematica} The equation can also be repeated with the initial density operator rotated by an angle $\chi$ about the $x$ or $y$ axes (because the $\hat{H}_{\pm}$ and $\hat{K}_{\pm}$ are rotationally invariant about the $z$ axis it does not matter which axis is chosen) to determine the dependence on the initial orientation of the polarisation relative to the molecular spin-orbit coupling axis. 
		
		The equations can also be solved for the glyceronitrile model where $k_\mathrm{f}=0$ and we instead introduce a spin-independent quenching rate $k_\mathrm{Q}$ by adding a term $(k_\mathrm{Q}/2)\hat{1}$ to both reaction operators $\hat{K}_{\pm}$. The resulting yield asymmetry is given in the following section.
		
		\vspace{-10pt}
		\section{Full expressions for yield asymmetry}\label{app-yields}
		\vspace{-10pt}
		The full expression for $f(J,\epsilon,k_\mathrm{f},k_\mathrm{r},\theta)$ is given by
		\begin{align}
			\begin{split}
				f&(J,\epsilon,k_\mathrm{f},k_\mathrm{r},\theta) = \frac{1}{8}\bigg(4 J^2 \left(32 k_{\mathrm{f}}^2+16 k_{\mathrm{f}} k_{\mathrm{r}}+k_{\mathrm{r}}^2\right)\\
				&+\cos (4 \theta ) \left(k_{\mathrm{f}} \left(4 \epsilon ^2 (4 k_{\mathrm{f}}+k_{\mathrm{r}})+k_{\mathrm{r}}^3\right)-4 J^2 k_{\mathrm{r}}^2\right)\\
				&+64 J k_{\mathrm{f}} \epsilon  \cos (2 \theta ) (2 k_{\mathrm{f}}+k_{\mathrm{r}})\\
				&+k_{\mathrm{f}} k_{\mathrm{r}}^2 (8 k_{\mathrm{f}}+3 k_{\mathrm{r}})+4 k_{\mathrm{f}} \epsilon ^2 (4 k_{\mathrm{f}}+3 k_{\mathrm{r}})\bigg).
			\end{split}
		\end{align}
		
		For the case where quenching is included, but $k_\mathrm{f}$ is set to zero we find the following $R/S$ yield asymmetry,
		\begin{align}
			\Delta\Phi &= \left(\frac{p_\mathrm{A}\coschi}{2}\right) \frac{4 J k_{\mathrm{Q}} k_{\mathrm{r}} \sin (2 \theta ) (2 k_{\mathrm{Q}}+k_{\mathrm{r}})}{g(J,\epsilon,k_\mathrm{b},k_\mathrm{r},\theta)}\\
			\begin{split}
				g&(J,\epsilon,k_\mathrm{b},k_\mathrm{r},\theta) =2 J^2 \left(8 k_{\mathrm{Q}}^2+8 k_{\mathrm{Q}} k_{\mathrm{r}}+k_{\mathrm{r}}^2\right)\\
				&-2 J^2 k_{\mathrm{r}}^2 \cos (4 \theta )+16 J k_{\mathrm{Q}} \epsilon  \cos (2 \theta ) (k_{\mathrm{Q}}+k_{\mathrm{r}})\\
				&+k_{\mathrm{Q}} (k_{\mathrm{Q}}+k_{\mathrm{r}}) \left((2 k_{\mathrm{Q}}+k_{\mathrm{r}})^2+4 \epsilon ^2\right).
			\end{split}
		\end{align}
		The enantiomeric excess of this reaction can also be found for full spin-polarisation, i.e. $p_A\cos\chi = 1$, and in the small $k_\mathrm{r}$ limit, this simplifies to
		\begin{align}
			\mathrm{ee} = \frac{2 J k_{\mathrm{Q}} \sin (2 \theta )}{4 J^2+4 J \epsilon  \cos (2 \theta )+k_{\mathrm{Q}}^2+\epsilon ^2} + \mathcal{O}(k_\mathrm{r}),
		\end{align}
		which is maximised when $k_{\mathrm{Q}} = \sqrt{4J^2+\epsilon^2+4J\epsilon\cos(2\theta)}$ and at
		\begin{align}
			\mathrm{ee}_\mathrm{max} = \frac{J\sin (2 \theta )}{\sqrt{\epsilon ^2+4 J \epsilon   \cos (2 \theta )+4J^2}}+ \mathcal{O}(k_\mathrm{r})
		\end{align}
		This maximum can exceed the $\sin(2\theta)$ limit if $\epsilon$ and $J$ have opposite signs. In fact it can in principle be maximised at 50\% independent of the magnitude of the spin-orbit coupling contribution to the electorn transfer.
		
		\section{Example radical pair reactions generating chiral products}

		\begin{figure}
			\centering
			\includegraphics[width=1.0\linewidth]{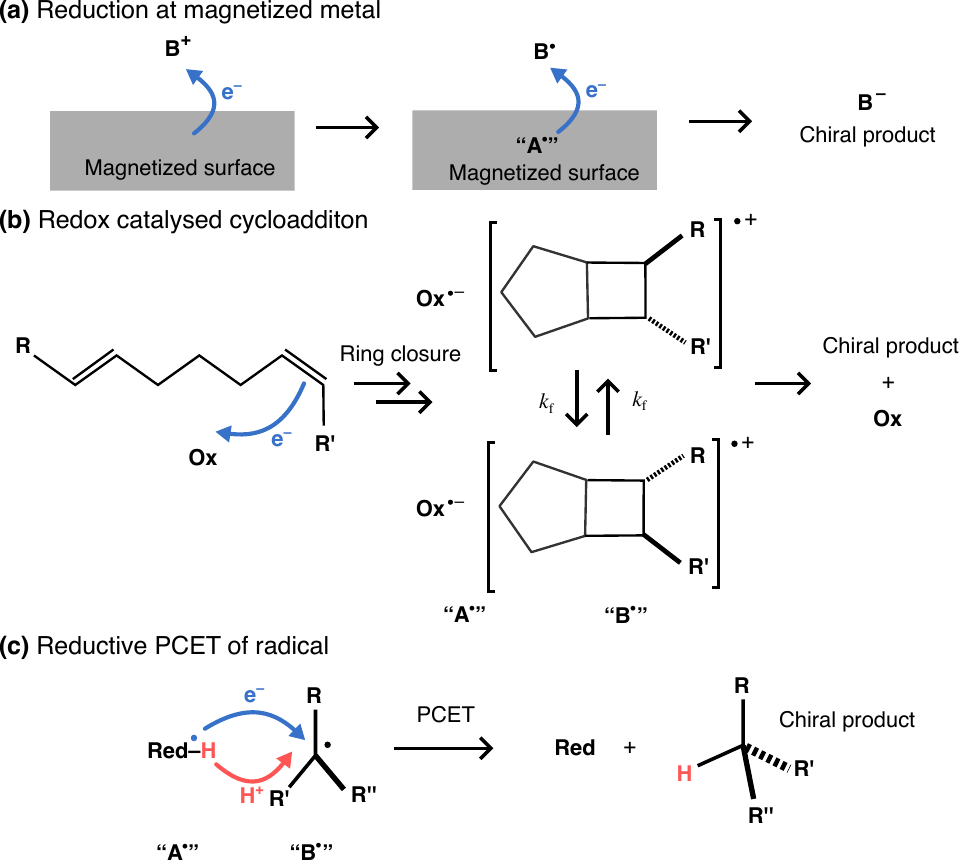}
			{\footnotesize\caption{Illustrations of potential systems which could realise the enantioselective radical pair mechanism. (a) Redox reactions at magnetized surfaces generating chiral products. (b) Redox catalysed cycloaddition reactions in which one of the intermediate radicals becomes spin-polarised. (c) PCET between an open-shell reducing agent and a prochiral carbon radical.}\label{fig-mechanisms}}
			\vspace{0pt}
		\end{figure}
		\hl{Let us now consider some other classes of reactions which could involved radical pair electron transfer, which we predict would be sensitive to electron spin polarization and chirality, if the intermediates are spin-polarized by a magnetized surafce acting as a redox catalyst.} A reaction in which stereochemistry is determined by the electron transfer reaction of a radical pair include the final step of (photo)redox catalysed cycloaddition reactions.\cite{metzger_dynamic_2021,morse_enantioselective_2018} If the catalysis were facilitated by redox at a magnetized solid surface, as in Fig.~\ref{fig-mechanisms}(a), then one of the radicals involved in the final redox step could become spin-polarized, leading to enantioselective synthesis of the final cycloaddition product. \hl{An example redox catalysed 2+2 cycloaddition is shown in Fig.~\ref{fig-mechanisms}(b) where a catalyst oxidises the reactant, forming a radical pair. The radical cation undergoes rapid ring-closure, potentially reversibly, so this radical pair exists in two chiral forms which can interconvert. The final irreversible step where the radical intermediate is quenched by the \ce{Ox^{$\bullet-$}} radical anion involves a spin-selective electron transfer between the radicals (see Scheme 2 in Ref.~\cite{morse_enantioselective_2018}), which must recombine to a closed shell state.\cite{morse_enantioselective_2018} It is therefore this last step which could be enantioselective if for example, and the radical cation could be formed in a spin-polarized state, if the initial oxidation occurs at a magnetized surface.
		} This is consistent with previously observed enantioselectivity of the 2+4 cycloaddition a diene and carbonyl at spin-polarized hematite nanoparticles; the small experimental enantiomeric excess \hl{of between 8.5\% and 16\%} is certainly within the predicted bound of 50\% for the mechanism proposed here.\cite{metzger_dynamic_2021} \hl{The other reaction where spin-polarization appeared to lead to enantioselective reactivity considered in Ref.~ \cite{metzger_dynamic_2021} is oxidation of an organic sulfide to sulfoxide on hematite nanoparticles. In the proposed mechanism for this is some iron containing systems does not involved radical intermediates, but some metal catalysed sulfide oxidations do involve sulfide radical cations, which would need to transfer another electron at some stage to form the final chiral sulfoxide. So there is a possibility that this reaction could also occur via radical pair intermediates, although evidence would point towards a different mechanism of enantioselectivity in this case.}
		
		As an additional example of a possible class of radical pair electron transfer reactions which could show enantioselectivity, proton-coupled electron transfer (PCET)\cite{brugh_anomalous_2020} or associative electron transfer (AET)\cite{zhang_single_2014} reaction of radicals or radical ions. This is illustrated in Fig.~\ref{fig-mechanisms}(c) for PCET of a prochiral carbon radical, where the final product is chiral, and although here we depict the electron donor radical \ce{Red^{$\bullet$}-H} to be both the electron and proton donor, this need not be the case. For this reaction electron withdrawing groups on the carbon radical could be used to promote PCET over direct hydrogen atom transfer. Alternatively associative electron transfer reactions at a prochiral centre, could also be enantioselective in the same way. \hl{As discussed above, these reactions would need to be catalysed by redox at a magnetized surface which can act as a source of spin-polarized electrons.}

		It should be noted that although in this work we have limited the discussion to redox reactions between pairs of radicals, it stands to reason that the same phenomenon could emerge in redox reactions of other open-shell systems with more unpaired electrons.\cite{steiner_magnetic_1989} This would mean that many redox reactions involving open-shell transition metal complexes could be enantioselective, or potentially reactions involving radical scavengers.\cite{kattnig_sensitivity_2017} The latter class of reactions have been shown theoretically to exhibit enhanced sensitivity to small perturbations in magnetic interactions in the radicals, so it stands to reason that these reactions could also show enhanced sensitivity to spin polarisation and spin-orbit interactions.\cite{kattnig_sensitivity_2017}
		
		\section{Experimental Verification}
		
		Having discussed some possible real-world examples of chemical reactions which could involve enantioselective radical pair electron transfers, it is worth discussing how one could experimentally verify this proposed mechanism. Clearly identification of radical intermediates, for example by (time-resolved) electron paramagnetic resonance (TR)EPR or transient absorption spectroscopy, would be an essential part of experimental verification, but this would not unambiguously distinguish this mechanism from others. Because this proposal involves the radical pair as an intermediate, and its transient coherent spin dynamics, then it should be sensitive to perturbations to these spin dynamics.\cite{steiner_magnetic_1989} For example sensitivity of enantiomeric excess to the magnitude of a static external magnetic field, or sensitivity to radio frequency magnetic fields,\cite{henbest_radio_2004} which would ``scramble'' electron spin coherence, would provide evidence for the radical pair mechanism of enantioselectivity. An additional possible experiment for confirming this mechanism would be through the introduction of magnetic isotopes, for example exchanging \ce{{}^{12}C} for \ce{{}^{13}C}, at prochiral centres.\cite{buchachenko_magnetic_2001} These magnetic isotopes could promote additional spin-relaxation and inhibit the enantioselectivity.
		
		As additional verification of the enantioselective radical pair mechanism, we note that all parameters in the model, namely electronic couplings and rate constants, can in principle be calculated using \textit{ab initio} electronic structure methods and molecular simulation (or more approximate theories). Furthermore effects of spin-relaxation and additional processes that affect radical pair spin dynamics can be modelled using established techniques from the theoretical framework of spin chemistry. In contrast, for the main other proposed mechanism for spin-polarisation producing enantioselective reactivity, it is unclear how to use existing tools in theoretical and computational chemistry to make quantitative predictions of enantioselectivity arising from this mechanism.

	\section*{References}
	\bibliography{refs}